\documentclass[aps,superscriptaddress,nofootinbib,eqsecnum,prd,notitlepage]{revtex4-2}

\pdfoutput=1

\usepackage{amsfonts}
\usepackage{amsmath}
\usepackage{amssymb}
\usepackage{graphicx,color}
\usepackage{float}
\usepackage{hyperref}
\usepackage{subfigure}
\usepackage{dcolumn}
\usepackage{soul}
\usepackage{ulem}
\usepackage{verbatim}
\usepackage{graphicx}
\usepackage{color}
\usepackage{xparse}

\begin{document}

\title{On the classicality and uniqueness in loop quantization of Bianchi-I spacetimes }

\author{Meysam Motaharfar}
\email{mmotah4@lsu.edu}
\affiliation{Department of Physics and Astronomy, Louisiana State University,
  Baton Rouge, LA 70803, USA}

  \author{Parampreet Singh}
\email{psingh@lsu.edu}
\affiliation{Department of Physics and Astronomy, Louisiana State University,
  Baton Rouge, LA 70803, USA}
\affiliation{Center for Computation and Technology, Louisiana State University,
  Baton Rouge, LA 70803, USA}

  \author{Eklavya Thareja}
\email{ethare1@lsu.edu}
\affiliation{Department of Physics and Astronomy, Louisiana State University,
  Baton Rouge, LA 70803, USA}

\begin{abstract}
In loop quantum cosmology, ambiguities in the Hamiltonian constraint can result in models with varying phenomenological predictions. In the homogeneous isotropic models, these ambiguities were settled, and the improved dynamics was found to be a unique and phenomenologically viable choice. This issue has remained unsettled on the inclusion of anisotropies, and in the Bianchi-I model there exist two generalizations of isotropic improved dynamics. In the first of these, labelled as $\bar \mu$ quantization, the edge length of holonomies depends on the inverse of the directional scale factor. This quantization has been favored since it results in universal bounds on energy density and anisotropic shear, and can be viably formulated for non-compact as well as compact spatial manifolds. However, there exists an earlier quantization, labelled as $\bar \mu'$ quantization, where edge lengths of holonomies depend on the inverse of the square root of directional triads. This quantization is also non-singular and so far believed to yield a consistent physical picture for spatially compact manifolds. We examine the issue of the physical viability of these quantizations for different types of matter in detail by performing a large number of numerical simulations. Our analysis reveals certain limitations which have so far remained unnoticed. We find that while being non-singular, the $\bar \mu'$ quantization suffers from a surprising problem where one of the triad components and associated polymerized term retains Planckian character even at large volumes. As a result, not only is the anisotropic shear not preserved across the bounce, which is most highlighted in the vacuum case, but the universe can exhibit an unexpected cyclic evolution. These problematic features are absent from the $\bar \mu$ quantization leaving it as the only viable prescription for loop quantizing the Bianchi-I model.

\end{abstract}

\maketitle


\section{Introduction}

The existence of singularities in Einstein's theory of general relativity (GR) \cite{Geroch:1968ut, Hawking:1970zqf, Borde:1993xh, Borde:2001nh} is one of the strongest motivations for the hunt of a theory of quantum gravity. It has been expected that a quantum theory of spacetime would result in a resolution of the big bang and black hole singularities. This expectation has been found to be true in the framework of loop quantum cosmology (LQC) where techniques of loop quantum gravity (LQG) have been applied to cosmological spacetimes \cite{Ashtekar:2011ni}. A key prediction of LQC is that the big bang singularity is replaced with a quantum bounce as spacetime curvature approaches the Planck regime \cite{Ashtekar:2006rx, Ashtekar:2006uz, Ashtekar:2006wn}. In fact, in the case of a spatially flat, homogeneous, and isotropic universe sourced with a massless scalar field, there exists a universal upper bound on the eigenvalues of the energy density operator determined by the area gap in the quantum geometry \cite{Ashtekar:2007em} and the probability for the occurrence of the bounce turns out to be unity in the consistent histories formulation of quantum mechanics \cite{Craig:2013mga}. The result of singularity resolution has been generalized to several spacetimes, including in the presence of spatial curvature \cite{Szulc:2006ep, Ashtekar:2006es,Vandersloot:2006ws, Szulc:2007uk}, inflationary potential \cite{Giesel:2020raf}, anisotropies \cite{Ashtekar:2009vc,Ashtekar:2009um,Wilson-Ewing:2010lkm} and in the presence of Fock quantized inhomogenities \cite{Garay:2010sk}.
Using an effective description of underlying quantum geometry, the phenomenological implications of LQC for various models have been extensively studied \cite{Agullo:2016tjh, Li:2021mop, Li:2023dwy}, including potential signatures in the CMB. Bounds on the energy density, expansion, and shear scalars in different models have been found \cite{Corichi:2008zb,Gupt:2011jh, Singh:2013ava}, and strong curvature singularities have been shown to be generically resolved in isotropic models \cite{Singh:2010qa, Saini:2018tto}, as well as anisotropic models \cite{Singh:2011gp,Saini:2017ipg,Saini:2017ggt,Saini:2016vgo}. These results in cosmological models have also been generalized to black hole spacetimes where the central singularity is resolved in the Planck regime \cite{Ashtekar:2023cod}.

Because of the underlying quantization ambiguities, there are distinct choices of quantum Hamiltonian constraint in LQC which can result in different phenomenological implications. An important exercise is to use various consistency requirements to restrict these choices and possibly find a unique choice. It turns out that in the case of isotropic universes, there indeed is a unique quantization, the so-called $\bar \mu$-scheme or the `` improved dynamics", which leads to a consistent infrared and ultraviolet behavior and is independent of problems associated with the rescaling of fiducial cell when one considers spatial non-compact manifolds \cite{Corichi:2008zb, Corichi:2009pp}. In the $\bar \mu$ scheme, when one equates the physical area of the loop with the minimum allowed area on the quantum geometry, one obtains $\bar \mu \propto |p|^{-1/2}$ where $p$ is the isotropic triad. An early attempt to generalize this to Bianchi-I spacetime replaced the isotropic triad with a directional triad $p_i$ in the expression of ``$\bar \mu_i$''. This approach was first developed by Chiou \cite{Chiou:2006qq}, later studied in more detail by Chiou and Vandersloot \cite{Chiou:2007sp} and thoroughly investigated by  Mena Marugan and collaborators \cite{Martin-Benito:2008dfr, Martin-Benito:2009xaf}. We refer to this quantization as the $\bar \mu'$ quantization in this paper. The resulting quantization shares similar features with an exactly solvable model in isotropic LQC but has limitations in the sense that there is no freedom in scaling the individual directions of the fiducial cell, leading to a restricted topology like a 3-torus \cite{Corichi:2009pp}. The energy density and anisotropic shear scalar are unbounded in $\bar \mu'$ quantization, and there is no universal quantum gravity scale for singularity resolution. However, for compact spatial manifolds with torus topology, the unboundedness of energy density and anisotropic shear can be addressed using the inverse triad modifications. Interestingly, Chiou and Vandersloot also discussed an alternative scheme (see appendix C of \cite{Chiou:2007sp}) where the edge lengths were chosen such that $\bar \mu_i \propto 1/a_i$ where $a_i$ denotes the directional scale factor. This quantization also agrees with $\bar \mu$-scheme in the isotropic limit and was developed in detail by Ashtekar and Wilson-Ewing \cite{Ashtekar:2009vc}. This prescription, hereafter referred to as the ``$\bar \mu$ quantization, successfully overcomes the limitation of restriction to compact manifolds and results in universal bounds on energy density and anisotropic shear \cite{Corichi:2009pp}. For these reasons, the $\bar \mu$ quantization is favored over the $\bar \mu'$ quantization in the literature, though the latter can be justified as a plausible alternative when one considers spatially-compact manifolds. Due to this reason, there have been two possible choices for loop quantization of Bianchi-I spacetime.

 In this manuscript, assuming the validity of effective description,  we take a closer look at the viability of these prescriptions guided by the requirement that any quantum theory of gravity to be viable must consistently recover the classical GR and its physical properties in the classical regime. Due to loop quantum effects, the singularity is resolved in both $\bar \mu$ and $\bar \mu'$ quantizations and the universe reaches a large volume compared to the Planck volume across the non-singular bounce. However, this is necessary but not sufficient to conclude that spacetime has become classical. In fact, spacetime only becomes classical when it also recovers all the physical properties of classical dynamics. Let us here note that an important physical observable in Bianchi-I is the anisotropic shear scalar, which measures the deviation from isotropic spacetime. The anisotropic shear is a constant of motion in the classical Bianchi-I spacetime, however, in loop quantum cosmological models, the big bang singularity is replaced with a quantum bounce, during which the anisotropic shear is not conserved due to quantum gravity effects. Such a bounce connects the expanding branch to the contracting branch before the big bang, whereby the large classical universe contracts, bounces back, and expands into the large classical universe. Therefore, if the universe is classical before and after the bounce at large volumes, one expects that the anisotropic shear scalar will be preserved across the bounce when it is compared at the same volume before and after the bounce in the absence of an anisotropic matter source such as  magnetic fields. Though this issue was  investigated in Ref. \cite{Chiou:2007sp} the conclusions were not clear. It was analytically demonstrated that {\it{if}} the values of triads are large before and after the bounce, the anisotropic shear is preserved across the bounce as in the classical Bianchi-I spacetime.  But the pertinent question is whether all directional triads necessarily become large across the bounce? And whether this implies that the quantum gravity modifications to the Hamiltonian constraint do not die even at macroscopic volumes? Note that in an anisotropic spacetimes since the physical volume $v = (p_1 p_2 p_3)^{1/2}$ it is quite possible for volume to take a macroscopic value with one of the triads taking a small value. In fact, in a classical Bianchi-I vacuum spacetime, this picture is generic. But whether or not this implies a breakdown of classicality in a particular quantization is far more clear. Due to the complicated form of Hamilton's equations, this is difficult to answer analytically, and one must perform numerical simulations to understand the behavior of the anisotropic shear across the bounce without making any additional assumptions on the nature of solutions.

 While earlier investigations seem to suggest that for the compact topologies, phenomenological differences in the $\bar \mu$ and the $\bar \mu'$ quantization is not important, we re-examine this issue in more detail for different types of matter employing a large number of numerical simulations using HPC for the cases of vacuum, massless scalar, dust, and radiation in the effective spacetime description. Investigating the time evolution of directional connections, we find that for the $\bar \mu$ quantization, if the universe starts from a large classical regime, it bounces and expands back to a large classical regime generically. The quantum gravity modifications to the classical Hamiltonian constraint are important only in the bounce regime and die quickly away from this regime.  However, in the case of $\bar \mu'$ quantization, starting from large values of triads in the pre-bounce regime, we find that one of the triads always remains in the quantum regime during the post-bounce evolution of the universe. This is a surprising result so far remaining elusive in previous investigations. While the post-bounce volume becomes large, the universe does not become classical even at large volumes because of one of the polymerized terms in the Hamiltonian constraint never becomes classical. Interestingly, in an anisotropic evolution, even in $\bar \mu$ scheme, one triad can be small after the bounce, but this happens in the regime when classical dynamics has been fully recovered.

Since in the $\bar \mu'$ scheme, the anisotropic universe remains in the quantum regime after the bounce, it is reasonable to suspect the conservation
of an anisotropic shear scalar across the bounce. Hence, comparing the value of the pre-bounce and post-bounce anisotropic shear scalar at a large volume regime while satisfying the Hamiltonian constraint with excellent accuracy, we find that the anisotropic shear is preserved across the bounce in the case of the $\bar \mu$ quantization for vacuum
spacetime and also including massless scalar fields, dust, and radiation matter components. However, we find that the anisotropic shear is not conserved across the bounce in the case of $\bar \mu'$ quantization, and it has a usually larger post-bounce value in comparison to the pre-bounce value. Moreover, we find that for particular initial conditions, the universe goes through an unexpected cyclic evolution in the case of adding dust or radiation, which is closely related to the quantum behavior of one of the triads. Our work shows that care needs to be exercised to understand the large volume behavior in anisotropic models, and one cannot take for granted the classicality condition in the effective dynamics of Bianchi-I LQC models. Our result implies that $\bar \mu'$ quantization does not yield a classical behavior on one of the two sides of the bounce and is thus not a viable scheme. This leaves the $\bar \mu$ quantization as a unique choice for the loop quantization of Bianchi-I spacetime.

The manuscript is structured as follows. In section \ref{section II} we provide a brief overview of the classical dynamics of Bianchi-I cosmology in terms of  symmetry-reduced Ashtekar-Barbero variables, namely triads and corresponding directional connections. We then establish their relationship to the familiar metric variables. In Section \ref{section III}, we present the loop quantization of the classical theory for both $\bar \mu$ quantization and $\bar \mu'$ quantizations at the level of effective dynamics and explore their physical properties. Section \ref{section IV} is dedicated to elaborating on the numerical methodology, presenting the numerical results for the classicality condition, and discussing the implications of not satisfying this condition in the case of $\bar \mu'$ quantization, including violation of the conservation of anisotropic shear scalar across the bounce and unexpected cyclic evolution of the universe. Note that in this section, we use Planck units. Finally, we give a summary of the results and conclusion in section \ref{section V}.


\section{Classical dynamics of Bianchi-I spacetime}\label{section II}

The loop quantization program is based on the classical gravitational phase space variables, namely the Ashtekar-Barbero connection $A_{a}^{i}$ and the triads $E_{i}^{a}$ (where $i=1, 2, 3$). Hence, it is useful to briefly review the dynamics of Bianchi-I spacetime, a spatially flat homogeneous (but anisotropic) universe, in a canonical framework and then relate it to the conventional metric variables. To this end, we consider the homogeneous (orthogonal) Bianchi-I spacetime with a spatial manifold $\mathbb{T}^3$ and lapse function $N=1$ given by
\begin{align}\label{Bianchi-I}
\mathrm{d}s^2 = - \mathrm{d}t^2 + a_{1}^2 \mathrm{d}x^2 + a_{2}^2 \mathrm{d}y^2 + a_{3}^2 \mathrm{d}z^2,
\end{align}
where $a_{1}$, $a_{2}$ and $a_{3}$ are directional scale factors, while the mean scale factor can be defined as $a:= (a_{1}a_{2}a_{3})^{1/3}$. Upon isotropization, i.e., when $a_{1}=a_{2}=a_{3}=a$, metric \ref{Bianchi-I} reduces to the Friedmann-Lemaître-Robertson-Walker (FLRW) metric describing a spatially flat, homogeneous, and isotropic universe. The Ashtekar-Barbero variables $A^{i}_{a}$ and $E_{i}^{a}$ reduce to connections $c^{i}$ and triads $p_{i}$ with only one independent component per spatial direction upon symmetry reduction and imposing the Gauss and the spatial-diffeomorphism constraints. The triads are kinematically related to the directional scale factors as follows
\begin{align}\label{triads}
|p_{1}| = a_{2}a_{3}, \ \ \ \ \ \ \ \ \ \ \ \ \ \ \ \ \ |p_{2}| = a_{1}a_{3}, \ \ \ \ \ \ \ \ \ \ \ \ \ \ \ \ \  |p_{3}| = a_{1}a_{2}.
\end{align}
The modulus sign arises because of the orientation of the triad. However, without losing generality, we here assume a positive sign for the orientation of directional triads. From these relations, one can find that directional scale factors are related to triads, i.e., $a_{1} = \sqrt{p_{2}p_{3}/p_{1}}$ (and similarly $a_{2}$ and $a_{3}$ with cyclic permutation). Moreover, in the phase space, the triads and connections satisfy the following Poisson bracket
\begin{align}
\{c^{i}, p_{j}\} = {8\pi G} \gamma \delta^{i}_{j},
\end{align}
where $\gamma \approx 0.2375$ is the Barbero-Immirzi parameter, which is fixed by black hole thermodynamics in LQG. The classical Hamiltonian constraint for matter content minimally coupled to the gravitational sector, in terms of directional connections $c_{i}$ and triads $p_{i}$, reads as
\begin{align}\label{Hamiltonian}
\mathcal{H}_{cl} = \mathcal{H}_{g} + \mathcal{H}_{m} = - \frac{1}{8 \pi G \gamma^2 v} \left(c_{1}p_{1} c_{2}p_{2} + c_{2}p_{2}c_{3}p_{3} + c_{3}p_{3}c_{1}p_{1}\right) + \mathcal{H}_{m},
\end{align}
where $\mathcal{H}_{m}$ is matter part of Hamiltonian and $v = \sqrt{p_{1}p_{2}p_{3}} = a_{1}a_{2}a_{3}$ is physical volume of a unit comoving cell \footnote{In non-compact models in LQC, one introduces a fiducial cell to define a symplectic structure whose coordinate lengths enter the relation between triads and scale factors. We assume the coordinate lengths of this fiducial cell to be unity.}. Given the classical Hamiltonian $\mathcal{H}_{cl}$, the dynamical equations for triads and connection components are determined using Hamilton's equations
\begin{align} \label{Hamilton's equations}
\dot p_{i} = \{p_{i}, \mathcal{H}_{cl}\} = - 8 \pi G \gamma \frac{\partial \mathcal{H}_{cl}}{\partial c_{i}}, \ \ \ \ \ \ \ \ \ \ \ \ \dot c_{i} = \{c_{i}, \mathcal{H}_{cl}\} = 8 \pi G \gamma \frac{\partial \mathcal{H}_{cl}}{\partial p_{i}},
\end{align}
where a dot denotes the derivative with respect to the cosmic time $t$. The first set of equations leads to $c_{i} = \gamma \dot a_{i} = \gamma H_{i} a_{i} $ which together with the Hamiltonian constraint $\mathcal{H}_{cl} \approx 0$ result in the following Friedmann equation for Bianchi-I spacetime
\begin{align}\label{Friedmann}
H_{1}H_{2}+ H_{2}H_{3} + H_{3}H_{1} &= 8 \pi G \rho,
\end{align}
where we considered a perfect fluid with a barotropic equation of state $\omega = P/\rho$ while $\rho = {\mathcal{H}_{m}}/{v}$ and $P = - \partial \mathcal{H}_{m}/\partial v$ are energy density and pressure of matter content, respectively. Moreover, $H_{i}$ denotes directional Hubble parameters, which are related to the time derivatives of triad components, such as
\begin{align}\label{directional-Hubble}
H_{1} \equiv \frac{\dot a_{1}}{a_{1}}= \frac{1}{2} \left(\frac{\dot p_{2}}{p_{2}} + \frac{\dot p_{3}}{p_{3}} - \frac{\dot p_{1}}{p_{1}}\right),
\end{align}
(and similarly for $H_{2}$ and $H_{3}$). Likewise, using the dynamical equations for directional connections, $c_{i}$, one can find the second Friedmann equation as follows
\begin{align}\label{Raychaudhuri}
H_{1}^2 + H_{2}^2 + H_{3}^2 + \dot H_{2} + \dot H_{3} &= - 8 \pi G P,
\end{align}
(and its cyclic permutation). One can check that Eqs. (\ref{Friedmann}) and (\ref{Raychaudhuri}) reduce to the first and second Friedmann equations for FLRW spacetime
\begin{align}\label{Friedmann-i}
H^2 = \frac{8\pi G}{3} \rho, \ \ \ \ \ \ \ \ \ \ \ \ \ \ \ \ \ \dot H = - 4\pi G (\rho + P),
\end{align}
in the limit $H_{1} = H_{2} = H_{3} = H = \frac{\dot a}{a}$ where $H$ is the mean Hubble parameter defined as
\begin{align}\label{mean-Hubble}
H \equiv \frac{1}{3}\left(H_{1} + H_{2} + H_{3}\right).
\end{align}
From Hamilton's equations (\ref{Hamilton's equations}) and after some algebra, it can be found that
\begin{align} \label{pij}
\frac{d}{dt}(p_{i}c_{i}) = 8 \pi G \gamma v^2 \left(\rho + p_{i} \frac{\partial \rho}{\partial p_{i}}\right),
\end{align}
assuming isotropic matter content, i.e., $p_{i} \frac{\partial \rho}{\partial p_{i}} = p_{j} \frac{\partial \rho}{\partial p_{j}}$, Eq. (\ref{pij}) thus yields
\begin{align}
\frac{d}{dt} \left(p_{i}c_{i} - p_{j}c_{j}\right) = 0,
\end{align}
which can be integrated to give
\begin{align}
p_{i}c_{i}-p_{j}c_{j} = \gamma \alpha_{ij},
\end{align}
with $\alpha_{ij}$ being a constant anti-symmetric $3 \times 3 $ matrix satisfying by construction $\alpha_{12} + \alpha_{23} + \alpha_{31}=0$ and factor of $\gamma$ is for convenience. Writing in terms of directional Hubble parameters, one obtains
\begin{align}\label{Hij}
H_{i}-H_{j} = \frac{\alpha_{ij}}{a_{1}a_{2}a_{3}} = \frac{\alpha_{ij}}{a^{3}}.
\end{align}
Given Eq. (\ref{Hij}), one can write generalized Friedmann equations which contain the information about the anisotropic shear. From Eq. (\ref{mean-Hubble}), one can find that
\begin{align}
    H^{2} = \frac{1}{3} \left(H_{1}H_{2} + H_{2}H_{3} + H_{3}H_{1}\right) + \frac{1}{18} \left [(H_{1}-H_{2})^{2} + (H_{2}-H_{3})^{2} + (H_{3}-H_{1})^{2}\right],
\end{align}
while the first term is related to energy density through Friedman equation (\ref{Friedmann}) and second term can be written in term of mean scale factor using Eq. (\ref{Hij}) as follow
\begin{align}
    H^2 = \frac{8 \pi G}{3}\rho + \frac{\Sigma^2}{a^{6}},
\end{align}
where $\Sigma^2 = (\alpha^2_{12} + \alpha^2_{23} + \alpha^2_{31})/18$ is a constant of motion, i.e., $\dot \Sigma = 0$. To investigate the formation of singularities and their structures in Bianchi-I spacetime, it is useful to define the expansion rate and the anisotropic shear scalar. The expansion rate is given by the trace of the expansion tensor
\begin{align}
\theta = H_{1}+ H_{2}+H_{3} = 3H,
\end{align}
while anisotropic shear scalar $\sigma^2$, which measures the deviation from isotropic spacetime, is a traceless part of the expansion tensor, which in terms of directional Hubble parameters $H_{i}$ is given by
\begin{align}\label{shear-tensor}
\sigma^2 = \sigma^{\mu\nu} \sigma_{\mu\nu} = \frac{1}{3} \left(\left(H_{1}- H_{2}\right)^2 + \left(H_{2}- H_{3}\right)^2 + \left(H_{3}- H_{1}\right)^2 \right).
\end{align}
Note that the above two relations are kinematical relations, so they are independent of whether the underlying theory
is GR or LQC. From Eq. (\ref{shear-tensor}), anisotropic shear scalar is related to $\Sigma$, i.e., $\sigma^2 = 6 \Sigma^2/a^6 $. Hence, one can derive generalized Friedmann equations as follows \cite{Tsagas:2007yx}:
\begin{align}\label{generalized}
H^2 = \frac{8 \pi G}{3} \rho + \frac{\sigma^2}{6}, \ \ \ \ \ \ \ \ \ \ \ \ \ \ \dot H = - 4\pi G \left(\rho + P\right) - \frac{ \sigma^2}{2}.
\end{align}
These equations reduce to the Friedmann equations given in (\ref{Friedmann-i}) for a flat, isotropic universe when $\sigma^2 =0$. Using above dynamical equations, one finds that at vanishing scale factors $\rho$, $\theta$ and $\sigma^2$ diverge, leading to the divergence of curvature invariants and the breakdown of geodesic evolution at the singularities. From the first generalized Friedmann equation together with the conservation law, i.e.,
\begin{align}
\dot \rho + 3H(\rho+ P)=0,
\end{align}
which implies $\rho \propto a^{-3(1+\omega)}$, one realizes that the anisotropic shear behaves as the energy density of a perfect fluid with the equation of state $\omega=1$ (similar to a massless scalar field). This means that the early universe could be isotropized in the presence of matter content with an equation of state $\omega < -1/3$ (such as a scalar field with plateau-like potential in inflationary scenarios \cite{Linde:1981mu, Guth:1980zm, Bardeen:1983qw}) in expanding universe and $\omega \gg 1$ (such as a scalar field with a negative potential in ekpyrotic scenarios \cite{Steinhardt:2002ih, Buchbinder:2007ad}) in a contracting universe. Moreover, the presence of anisotropy makes the structure of singularities much richer. Although in an isotropic universe the big bang singularity is always point-like, in Bianchi-I spacetime the geometrical nature of the singularity depends on whether all three directional scale factors approach zero, leading to different types of singularities, namely, point-like, cigar-like, barrel-like, and pancake singularities. However, the point-like and cigar-like singularities are the most prevalent ones. In fact, if the energy density term dominates over the anisotropic term in the first equation of (\ref{generalized}) near the singularity, the approach to the singularity can be point-like, otherwise, it will be cigar-like (see \cite{Gupt:2012vi} for more discussion).


\section{Effective dynamics of Bianchi-I mdoel in LQC: $\bar \mu$ vs $\bar \mu'$ quantization}\label{section III}

Despite the fact that the complete theory of LQG has yet to be developed, its techniques can still be employed to examine symmetry-reduced spacetimes. In practice, LQC utilizes methods and concepts from LQG to perform quantization of cosmological spacetimes with reduced symmetries. To quantize, the classical Hamiltonian constraint is formulated using the fundamental variables of quantum theory, which are the holonomies of the connections evaluated along closed loops and the fluxes of the triads (which are directly proportional to the triads). Consequently, a discrete quantum difference equation arises, which governs the evolution of the universe. Interestingly, the quantum difference equation is non-singular and leads to the continuous Wheeler-DeWitt equation when the spacetime curvature becomes small. However, it was demonstrated that the underlying quantum dynamics of various cosmological spacetimes, such as the isotropic model \cite{Diener:2013uka, Diener:2014mia, Diener:2014hba} and the Bianchi-I model \cite{Diener:2017lde, Martin-Benito:2008dfr, Martin-Benito:2009xaf} can be accurately captured by a continuum effective description under some reasonable assumptions. In fact, the effective Hamiltonian for Bianchi-I LQC can be obtained by replacing the classical directional connections $c_{i}$ with bounded trigonometric functions, i.e., $c_{i} \rightarrow \sin(\bar \mu_{i} c_{i})/\bar \mu_{i}$, therefore,
\begin{align}
\mathcal{H} = - \frac{1}{8 \pi G \gamma^2 v} \left(\frac{\sin (\bar \mu_{1} c_{1})}{\bar \mu_{1}} \frac{\sin (\bar \mu_{2} c_{2})}{\bar \mu_{2}} p_{1}p_{2} + \textrm{cyclic permutations}\right) + \mathcal{H}_{m},
\end{align}
where $\bar \mu_{i}$ are real functions of triads (which are assumed to have positive orientation) that measure the discreteness of spacetime, and Hamiltonian constraint reduces to classical Hamiltonian constraint in the limit $\bar \mu_{i} c_{i}\ll 1$ (more specifically, when $\bar \mu_{i}c_{i} \sim n\pi$). Hereafter we call this condition the ``classicality condition''. In the early development of LQC, $\bar \mu_{i}$ was a constant, which was called the $ \mu_{0}$ scheme, however, later it was shown that such a scheme cannot recover classical GR at large volumes in the case of an isotropic LQC model \cite{Ashtekar:2006rx}. Then, it was argued that $\mu_{0}$ should proportional to inverse of scale factor, i.e., $\bar \mu \propto 1/\sqrt{|p|}$ or equivalently  $\bar \mu \propto 1/a$ in the isotropic spacetime, leading into the so-called $\bar \mu$ scheme or ``improved dynamics". To extend the $\bar \mu$ scheme to Bianchi-I spacetime with three triads, there is some ambiguity, which has resulted in two different quantizations so far: $\bar \mu'$ and $\bar \mu$ quantizations.

\subsection{The $\bar \mu'$ scheme}

In the case of $\bar \mu'$, the holonomy edge lengths depend on just one of the triads, therefore,
\begin{align}
\bar \mu_{1}' = \lambda \sqrt{\frac{1}{p_{1}}}, \ \ \ \ \ \ \ \ \ \ \ \bar \mu_{2}' = \lambda \sqrt{\frac{1}{p_{2}}}, \ \ \ \ \ \ \ \ \ \ \ \bar \mu_{3}' = \lambda \sqrt{\frac{1}{p_3}},
\end{align}
where $\lambda^2 = \Delta = 4 \sqrt{3}\pi \gamma \l_{Pl}^2$ with $\Delta$ being the minimum eigenvalue of the area operator in LQG. This model reduces to its isotropic model, i.e., $ \bar \mu_{1}' = \bar \mu_{2}' = \bar\mu_{3}' = \bar \mu = \lambda/\sqrt{p}$ when $p_{1} = p_{2} = p_{3} = p$. Given the holonomy edge lengths, the effective Hamiltonian for $\bar \mu'$ quantization can be rewritten as
\begin{align}
\mathcal{H}^{(\bar \mu')} = - \frac{1}{8 \pi G \gamma^2 \lambda^2} \left({\sin (\bar \mu_{1}' c_{1})} {\sin (\bar \mu_{2}' c_{2})}\frac{p_{1}p_{2}}{\sqrt{p_{3}}}+ \textrm{cyclic permutations}\right) + \mathcal{H}_{m}.
\end{align}
With this effective Hamiltonian, one can find the dynamical equations via Hamilton's equations, as in classical spacetime. Resulting Hamilton's equations for $\bar \mu'$ quantization read as
\begin{align}\label{triad-Madrid}
\dot p_{1} = \frac{p_{1}}{\gamma \lambda v} \cos(\bar \mu_{1}' c_{1})\left(p_{2}^{\frac{3}{2}} \sin(\bar \mu_{2}' c_{2}) + p_{3}^{\frac{3}{2}} \sin(\bar \mu_{3}'c_{3})\right) ,
\end{align}
\begin{align}\label{connection-Madrid}
\nonumber \dot c_{1} &= - \frac{1}{\gamma \lambda^2}\left [ {\sin (\mu_{1}' c_{1})} {\sin (\mu_{2}' c_{2})} \frac{p_{2}}{\sqrt{p_{3}}} -\frac{1}{2} {\sin (\mu_{2}' c_{2})} {\sin (\mu_{3}' c_{3})} \frac{p_{2}p_{3}}{p_{1}^{\frac{3}{2}}} + {\sin (\mu_{3}' c_{3})} {\sin (\mu_{1}' c_{1})} \frac{p_{3}}{\sqrt{p_{2}}} \right] \\ &+ \frac{c_1}{2\gamma \lambda v} \cos(\mu_{1}' c_{1}) \left[p_{2}^{\frac{3}{2}}\sin(\mu_{2}' c_{2}) + p_{3}^{\frac{3}{2}}\sin(\mu_{3}' c_{3})\right] + 8 \pi G \frac{\partial \mathcal{H}_{m}}{\partial p_{1}}.
\end{align}
The other four equations (two for triads and two for connections) can easily be derived by cyclic permutation of Eqs. (\ref{triad-Madrid}) and (\ref{connection-Madrid}). From the Hamiltonian constraint, i.e., $\mathcal{H}^{(\bar \mu')} \approx 0$, the energy density is given by
\begin{align}\label{rho-Madrid}
\rho = \frac{1}{8 \pi G \gamma^2 \lambda^2} \left({\sin (\bar \mu_{1}' c_{1})} {\sin (\bar \mu_{2}' c_{2})} \frac{\sqrt{p_{1}p_{2}}}{p_{3}} + {\sin (\bar \mu_{2}' c_{2})} {\sin (\bar \mu_{3}' c_{3})}\frac{\sqrt{p_{2}p_{3}}}{p_1} + {\sin (\bar \mu_{3}' c_{3})} {\sin (\bar \mu_{1}' c_{1})}\frac{\sqrt{p_{3}p_{1}}}{p_2}\right),
\end{align}
where we used the fact that $\mathcal{H}_{m} = \rho v$. From the expression for energy density, one realizes that there are some terms that are proportional to triads, and hence they are unbounded. This means that the $\bar \mu'$ quantization does not result in a universal quantum gravity scale, as is the case in isotropic spacetime (see Ref. \cite{Corichi:2009pp} for a comprehensive discussion about this issue). Using Hamilton's equations for triads and Eq. (\ref{directional-Hubble}), the directional Hubble parameters are given by
\begin{align}
\nonumber H_{1} &= \frac{1}{2 \gamma \lambda v} \left(p_{1}^{\frac{3}{2}} \sin(\bar \mu_{1}' c_{1}) (\cos(\bar \mu_{2}' c_{2}) + \cos(\bar \mu_{3}' c_{3})) + p_{2}^{\frac{3}{2}} \sin(\bar \mu_{2}' c_{2}) (\cos(\bar \mu_{3}' c_{3}) - \cos(\bar \mu_{1}' c_{1})) \right. \\ & \left. + p_{3}^{\frac{3}{2}} \sin(\bar \mu_{3}' c_{3}) (\cos(\bar \mu_{2}' c_{2}) - \cos(\bar \mu_{1}' c_{1}))\right),
\end{align}
and similar equations can be derived for $H_{1}$ and $H_{2}$ using cyclic permutation. Using Hamilton's equations for triads, the expansion rate reads as
\begin{align}
\nonumber \theta &= \frac{1}{2\gamma \lambda v} \Bigg(p_{1}^{\frac{3}{2}} \sin (\bar \mu_{1}' c_{1}) \Big(\cos(\bar \mu_{2}' c_{2}) + \cos(\bar \mu_{3}' c_3)\Big) + p_{2}^{\frac{3}{2}} \sin (\bar \mu_{2}' c_{2}) \Big(\cos(\bar \mu_{1}' c_{1}) + \cos(\bar \mu_{3}' c_3)\Big) \
\\ &+ p_{3}^{\frac{3}{2}} \sin (\bar \mu_{3}' c_{3}) \Big(\cos(\bar \mu_{1}' c_{1}) + \cos(\bar \mu_{2}' c_2)\Big)\Bigg),
\end{align}
while having the directional Hubble parameter, one can also find the shear scalar using Eq. \ref{shear-tensor} as follows
\begin{align}\label{shear-Madrid}
\nonumber \sigma^{2} &= \frac{1}{3\gamma^2\lambda^2 v^2} \left[\left(\cos(\bar \mu_{1}'c_{1}) (p_{2}^{\frac{3}{2}}\sin( \bar \mu_{2}' c_{2}) + p_{3}^{\frac{3}{2}}\sin(\bar \mu_{3}' c_{3})) - \cos(\bar \mu_{2}' c_{2}) ( p^{\frac{3}{2}}_{3}\sin( \bar \mu_{3}' c_{3}) + p_{1}^{\frac{3}{2}} \sin(\bar \mu_{1}' c_{1}))\right)^2 \right. \\ & \nonumber \left. + \left(\cos(\bar \mu_{2}' c_{2}) (p_{3}^{\frac{3}{2}}\sin(\bar \mu_{3}' c_{3}) + p_{1}^{\frac{3}{2}} \sin(\bar \mu_{1}' c_{1})) - \cos(\bar \mu_{3}' c_{3}) (p_{1}^{\frac{3}{2}}\sin(\bar \mu_{1}' c_{1}) + p_{2}^{\frac{3}{2}}\sin(\bar \mu_{2}' c_{2}))\right)^2 \right. \\ & \left. + \left(\cos(\bar \mu_{3}' c_{3}) (p_{1}^{\frac{3}{2}}\sin(\bar \mu_{1}' c_{1}) + p_{2}^{\frac{3}{2}}\sin(\bar \mu_{2}' c_{2})) - \cos(\bar \mu_{1}' c_{1}) (p_{2}^{\frac{3}{2}}\sin( \bar \mu_{2}' c_{2}) + p_{3}^{\frac{3}{2}}\sin(\bar \mu_{3}' c_{3}))\right)^2 \right].
\end{align}
One can see from the expressions for expansion rate and shear scalar that there are some terms proportional to triads that are not bounded, meaning that expansion rate and shear scalar are also unbounded, similar to the energy density. However, one should note that the discreteness of spacetime also leads to other modifications in the Plack regime in LQC. The most important one is the inverse scale factor modification, which comes from the requirement of having a well-defined operator for the inverse of some power of scale factor, which in its spectra contains the zero eigenvalue. We can see from the expressions for expansion rate and shear anisotropic scalar that there are some inverse volume prefactors that will be modified upon applying the inverse scale factor correction. Hence, one can argue that taking into account the inverse scale factor modifications in the case of $\bar \mu'$ quantization may result in a bounded energy density and an anisotropic shear scalar.\footnote{The inverse scale factor modifications can result in non-trivial modifications in certain situations. See for eg. \cite{Motaharfar:2022pjp, Motaharfar:2023gpp} where they play an important role for initial conditions of the universe and the tunneling wavefunction proposal.}

\subsection{The $\bar \mu$ scheme}
The unboundedness of energy density and shear anisotropy in the absence of inverse scale factor modification for the $\bar \mu'$ scheme was one of the motivations which led to the development of another quantization -- the $\bar \mu$ scheme  which results in a universal quantum gravity scale for the observable quantities for both compact and non-compact manifolds. In fact, it was realized that one can also use loop quantization techniques if the holonomy edge lengths depend on inverse of directional scale factors similar to isotropic LQC \cite{Ashtekar:2009vc} resulting in the $\bar \mu$ quantization of Bianchi-I spacetime. In fact, the holonomy edge lengths in $\bar \mu$ quantization take the following form
\begin{align}
\bar \mu_{1} = \lambda \sqrt{\frac{p_{1}}{p_{2}p_{3}}},\ \ \ \ \ \ \ \ \ \ \ \bar \mu_{2} = \lambda \sqrt{\frac{p_{2}}{p_{1}p_{3}}}, \ \ \ \ \ \ \ \ \ \ \ \bar \mu_{3} = \lambda \sqrt{\frac{p_{3}}{p_{1}p_{2}}}.
\end{align}
Hence, the effective Hamiltonian for $\bar \mu$ quantization can be rewritten as follows
\begin{align}
\mathcal{H}^{(\bar \mu)} = - \frac{v}{8 \pi G \gamma^2 \lambda^2} \left({\sin (\bar \mu_{1} c_{1})} {\sin (\bar \mu_{2} c_{2})}+ \textrm{cyclic permutations}\right) + \mathcal{H}_{m},
\end{align}
from which one can derive the corresponding Hamilton's equations for $\bar \mu$ quantization given by
\begin{align}\label{triad-AWE}
\dot p_{1} = \frac{p_{1}}{\gamma \lambda} \cos(\bar \mu_{1} c_{1}) \left(\sin(\bar \mu_{2} c_{2}) + \sin(\bar \mu_{3}c_{3})\right) ,
\end{align}
\begin{align}\label{connection-AWE}
\nonumber \dot c_{1} &= 8 \pi G \gamma \frac{\partial \mathcal{H}_{m}}{\partial p_{1}} -\frac{v}{2\gamma \lambda^2
p_{1}} \left[ \bar \mu_{1} c
_{1} \cos(\bar\mu_{1}c_{1}) \left(\sin(\bar \mu_{2}c_{2}) + \sin(\bar\mu_{3}c_{3})\right) - \bar \mu_{3} c
_{3} \cos(\bar\mu_{3}c_{3}) \left(\sin(\bar \mu_{1}c_{1}) + \sin(\bar\mu_{2}c_{2})\right) \right. \\ & \left. - \bar \mu_{2} c
_{2} \cos(\bar\mu_{2}c_{2}) \left(\sin(\bar \mu_{3}c_{3}) + \sin(\bar\mu_{1}c_{1})\right) + \left(\sin(\bar\mu_{1}c_{1}) \sin(\bar \mu_{2}c_{2}) + \sin(\bar\mu_{2}c_{2} ) \sin(\bar \mu_{3}c_{3}) + \sin(\bar \mu_{3}c_{3} ) \sin(\mu_{1}c_{1})\right)\right] .
\end{align}
Similarly, the other four equations for triads and connections can be obtained by a cyclic permutation of Eqs. (\ref{triad-AWE}) and (\ref{connection-AWE}). Using Hamiltonian constraints, the expression for energy density reads as
\begin{align}
 \rho = \frac{1}{8\pi G \gamma^2 \lambda^2} \left({\sin (\bar \mu_{1} c_{1})} {\sin (\bar \mu_{2} c_{2})} + {\sin (\bar \mu_{2} c_{2})} {\sin (\bar \mu_{3} c_{3})} + {\sin (\bar \mu_{3} c_{3})} {\sin (\bar \mu_{1} c_{1})}\right).
\end{align}
One realizes that although the energy density diverges in the classical GR near the singularities, it is universally bounded in LQC in the case of $\bar \mu$ quantization, i.e.,
\begin{align}
\rho \leq \rho_{\textrm{max}} = \frac{3}{8 \pi G \gamma^2 \lambda^2} \simeq 0.41 \rho_{Pl},
\end{align}
where $\rho_{\textrm{max}}$ is the maximum energy for which the bounce occurs. Now, using Eq. (\ref{directional-Hubble}) and Hamilton's equations for triads, one can find the directional Hubble parameter as
\begin{align}
\nonumber H_{1} &= \frac{1}{2 \gamma \lambda} \Big( \sin(\bar \mu_{1} c_{1}) (\cos(\bar \mu_{2}c_{2}) + \cos(\bar \mu_{3}c_{3})) + \sin(\bar \mu_{2} c_{2}) (\cos(\bar \mu_{3}c_{3}) - \cos(\bar \mu_{1}c_{1})) \\ & + \sin(\bar \mu_{3} c_{3}) (\cos(\bar \mu_{2}c_{2}) - \cos(\bar \mu_{1}c_{1}))\Big),
\end{align}
and similar equations can be found for $H_{2}$ and $H_{3}$ by cyclic permutation. Using Hamilton's equations for $\bar \mu$ quantization, the expansion rate can be written as follows:
\begin{align}
\nonumber \theta &= \frac{1}{2\gamma \lambda} \Bigg(\sin (\bar \mu_{1}c_{1}) \Big(\cos(\bar \mu_{2}c_{2}) + \cos(\bar \mu_{3}c_3)\Big) + \sin (\bar \mu_{2}c_{2}) \Big(\cos(\bar \mu_{1}c_{1}) + \cos(\bar \mu_{3}c_3)\Big) \
\\ &+ \sin (\bar \mu_{3}c_{3}) \Big(\cos(\bar \mu_{1}c_{1}) + \cos(\bar \mu_{2}c_2)\Big)\Bigg),
\end{align}
and using directional Hubble parameters and Eq. (\ref{shear-tensor}), one reaches the following equation for a anisotropic shear scalar
\begin{align}\label{shear-AWE}
\nonumber \sigma^2 & = \frac{1}{3\gamma^2\lambda^2} \left[\left(\cos(\bar \mu_{1}c_{1}) (\sin( \bar \mu_{2} c_{2}) + \sin(\bar \mu_{3}c_{3})) - \cos(\bar \mu_{2}c_{2}) (\sin( \bar \mu_{3} c_{3}) + \sin(\bar \mu_{1}c_{1}))\right)^2 \right. \\ & \nonumber \left. + \left(\cos(\bar \mu_{2}c_{2}) (\sin(\bar \mu_{3} c_{3}) + \sin(\bar \mu_{1}c_{1})) - \cos(\bar \mu_{3}c_{3}) (\sin(\bar \mu_{1} c_{1}) + \sin(\bar \mu_{2}c_{2}))\right)^2 \right. \\ & \left. + \left(\cos(\bar \mu_{3}c_{3}) (\sin(\bar \mu_{1} c_{1}) + \sin(\bar \mu_{2}c_{2})) - \cos(\bar \mu_{1}c_{1}) (\sin( \bar \mu_{2} c_{2}) + \sin(\bar \mu_{3}c_{3}))\right)^2 \right].
\end{align}
Unlike the classical theory, the shear scalar is universally bounded in LQC for $\bar \mu$ quantization. The upper bound on shear scalar is given by
\begin{align}
\sigma^2 \leq \sigma^2_{\textrm{max}} = \frac{10.125}{3\gamma^2 \lambda^2} \simeq 11.57 l_{\mathrm{Pl}}^{-2}.
\end{align}
The boundedness of energy density, the expansion rate, and the shear scalar strongly indicate that singularities are avoided in Bianchi-I LQC. In fact, it has been demonstrated that, for arbitrary matter, the effective spacetime of the Bianchi-I model in LQC does not contain strong curvature singularities \cite{Singh:2011gp}. Additionally, numerical simulations indicate that these singularities are replaced by a bounce, and in the context of Bianchi-I spacetimes, this bounce is accompanied by Kasner transitions in the geometry of the spacetime \cite{Gupt:2012vi}. Although the existence of bounce is a generic feature of effective Bianchi-I spacetimes similar to their isotropic counterparts, there are significant differences due to the presence of anisotropies. In contrast to the isotropic case, the relationship between the energy density and anisotropic shear in the Bianchi-I model can lead to neither $\rho$ nor $\sigma^2$ reaching their maximum values, $\rho_{\textrm{max}}$ and $\sigma_{\textrm{max}}^2$, at the bounce. Due to the very complicated expressions for dynamical equations, finding a modified generalized Friedmann equation containing loop quantum gravity effects is not possible. However, exhaustive numerical analysis for different matter types in Ref. \cite{McNamara:2022dmf} showed that there is a parabolic relation between energy density at the bounce and the value of anisotropic shear scalar at the bounce.

To summarize this section, let us note that although these two quantizations are quite different on the theoretical grounds, it has long been believed that both of them result in the same physical properties at large volume regimes where quantum effects are negligible. In fact, one expects that the quantum universe will become classical at large volumes if the classicality condition is satisfied. Therefore, as it was discussed in Ref. \cite{Chiou:2007sp}, although the anisotropic shear is not a constant of motion during the quantum bounce phase, it is conserved when it is compared at the large volume regime before and after the bounce. However, as we will discuss in the next chapter, this is not always the case. The goal of the next section is to carry out an exhaustive numerical investigation to check whether Bianchi-I loop quantum cosmological models become classical.


\section{Numerical analysis}\label{section IV}

\begin{figure}[tbh!]
    \centering
    \includegraphics[scale= 0.72]{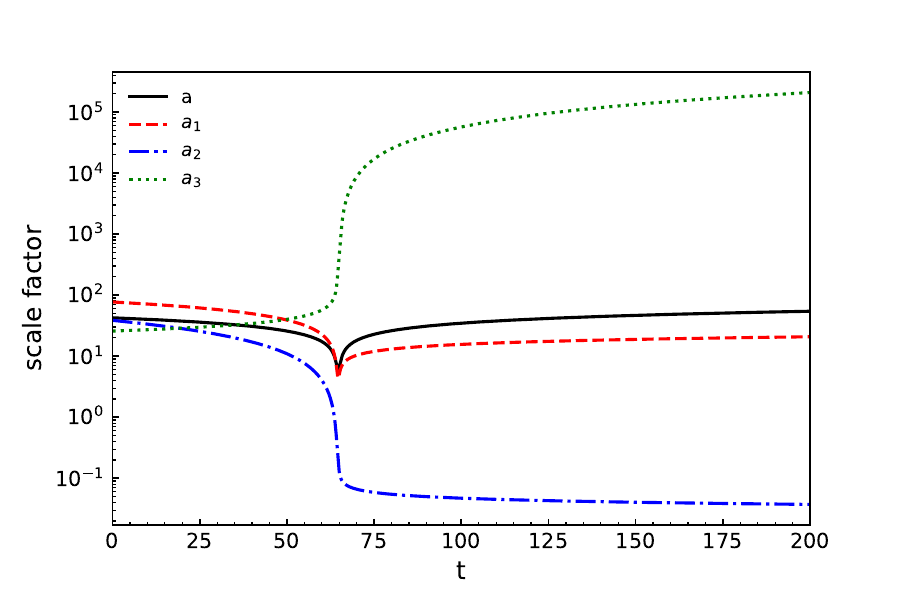}
    \caption{Evolution of directional scale factors and mean scale factor versus time with $p_{1}=1000$, $p_{2}=2000$, $p_{3}=3000$, $c_{1}= -0.13$, $c_{2} = -0.12$ for vacuum Bianchi-I spacetime in the case of $\bar \mu$ quantization. The big bang singularity is resolved and the universe becomes classical before and after the bounce.}
    \label{fig1}
\end{figure}

\begin{figure}[t!]
    \centering
    \includegraphics[scale= 0.6]{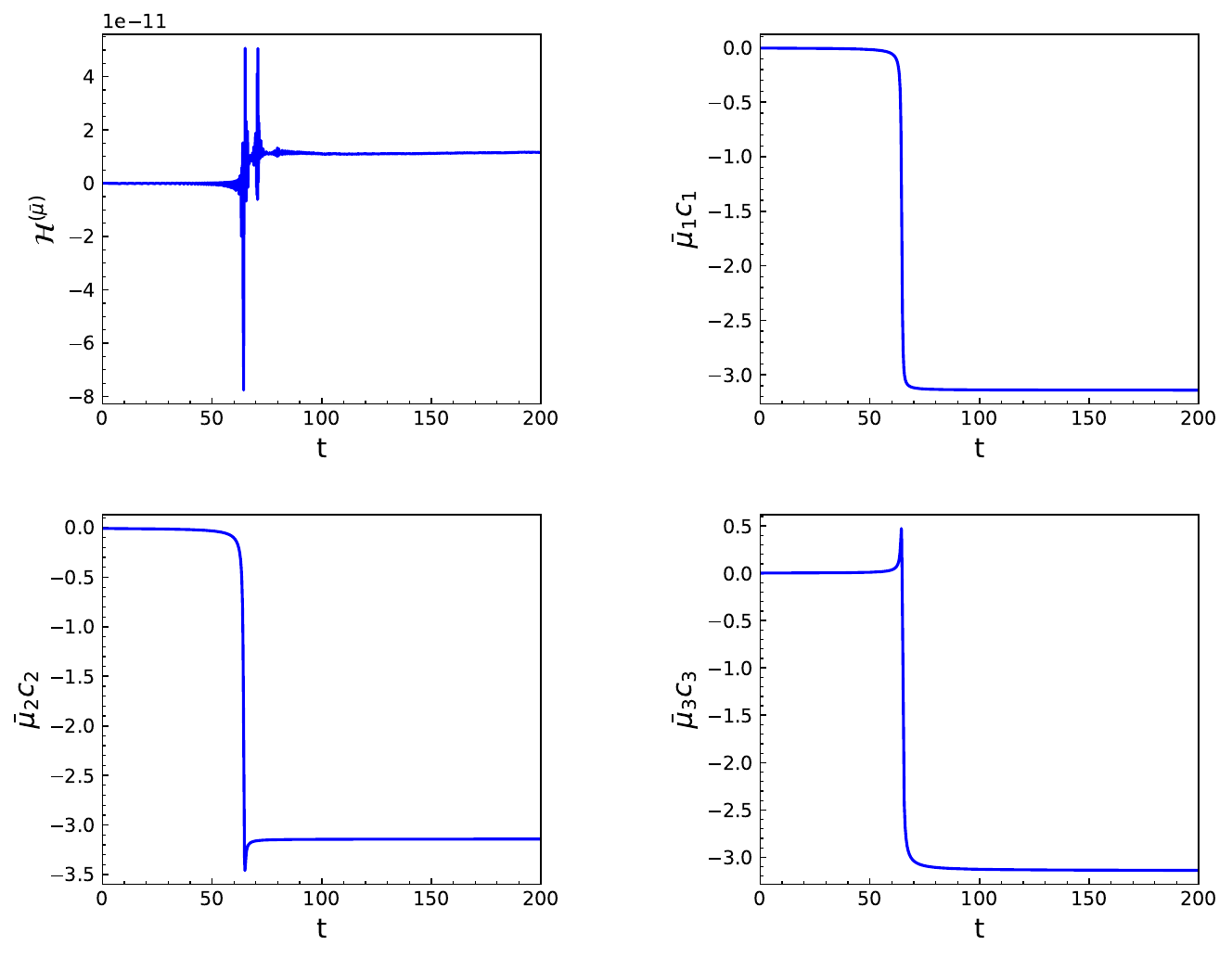}
    \caption{Evolution of effective Hamiltonian constraint and $\bar\mu_{i}c_{i}$ versus time with $p_{1}=1000$, $p_{2} =2000$, $p_{3}=3000$, $c_{1}= -0.13$, and $c_{2}= -0.12$ for vacuum Bianchi-I spacetime in the case of $\bar \mu$ quantization. As can be interpreted from this figure the polymerized corrections are negligible before and after the bounce. }
    \label{fig2}
\end{figure}

Due to the occurrence of the bounce, the universe can be extended to the contracting branch before the big bang in loop quantum cosmological models. This implies that the universe may start in a large classical regime, contract, reach a quantum bounce turnaround point, and then expand into a large classical regime. Hence, one expects to recover the physical properties of classical Bianchi-I spacetime at large volume regimes in both pre-bounce and post-bounce branches. As we discussed in section \ref{section II}, one of the most important properties of classical Bianchi-I spacetime is the conservation of anisotropic shear, assuming isotropic matter content. Although, this conservation breaks down during the quantum evolution of the bounce phase, as it is clear from Eqs. (\ref{shear-Madrid}) for $\bar \mu'$ quantization and (\ref{shear-AWE}) for $\bar \mu$ quantization, one expects to recover such conservation by comparing anisotropic shear before and after the bounce at large volumes {\it{if}} the universe becomes classical and effective spacetime is approximated by classical Bianchi-I spacetime. In fact, if one assumes the classicality conditions, the effective dynamics equations reduce to Friedmann equations for classical Bianchi-I spacetime with conserved anisotropic shear \cite{Chiou:2007sp}. As we will notice, although one can consider this assumption kinematically, the dynamics of the system does not always allow the universe to become classical, satisfying classicality conditions as it is in the case of $\bar \mu'$ quantization. Hence, as we will discuss below, such quantum behavior not only results in the violation of anisotropic shear conservation at large volume regimes but also induces some peculiar behavior, such as an unexpected cyclic evolution of the universe in the case of $\bar \mu'$ quantization.

\subsection{Classicality condition}

The first consistency check for any loop quantum cosmological model is whether it recovers the classical GR. In the Bianchi-I LQC model, one expects that Hamilton's Eqs. (\ref{triad-Madrid}) and \ref{connection-Madrid} for $\bar \mu'$ quantization and Eqs. (\ref{triad-AWE}) and (\ref{connection-AWE}) for $\bar \mu$ quantization reduce into the corresponding Friedmann Eqs. (\ref{Friedmann}) and (\ref{Raychaudhuri}) for classical Bianchi-I spacetime once the classicality condition, i.e., $|\bar \mu_{i}c_{i}|\ll 1$, is satisfied. To confirm that, we fix the initial values of triads in the contracting branch to be at a comparably large volume, then we determine one of the directional connections (in this case $c_{3}$) using the Hamiltonian constraint, i.e., $\mathcal{H}^{(\bar \mu)} =0$ for $\bar \mu$ qunatization, and $\mathcal{H}^{(\bar \mu')} =0$ for $\bar \mu'$ quantization, and fix the other two connection components while forcing all three directional connections to satisfy classicality conditions. To solve the six dynamical equations given in Eqs. (\ref{triad-Madrid}) and (\ref{connection-Madrid}) for $\bar \mu'$ quantization, and Eqs. (\ref{triad-AWE}) and (\ref{connection-AWE}) for $\bar \mu$ quantization, respectively, we use \textsc{numbalsoda} package which is for solving ordinary differential equation initial value problems, and then we apply the \textsc{dop853} algorithm, which is an explicit Runge-Kutta of order $8(5, 3)$ by Dormand and Prince with adaptive step size control. The advantage of this algorithm is that it uses a higher order correction to adapt the step size, and as such, it controls the error propagating in the system of equations. Moreover, one can change the values of absolute and relative tolerances to reduce the error in the Hamiltonian constraint. We choose the relative and absolute tolerances to be $10^{-14}$ and $10^{-11}$, respectively. However, we adjust their values depending on the initial conditions if higher accuracy is needed.

\begin{figure}[t!]
    \centering
    \includegraphics[scale= 0.72]{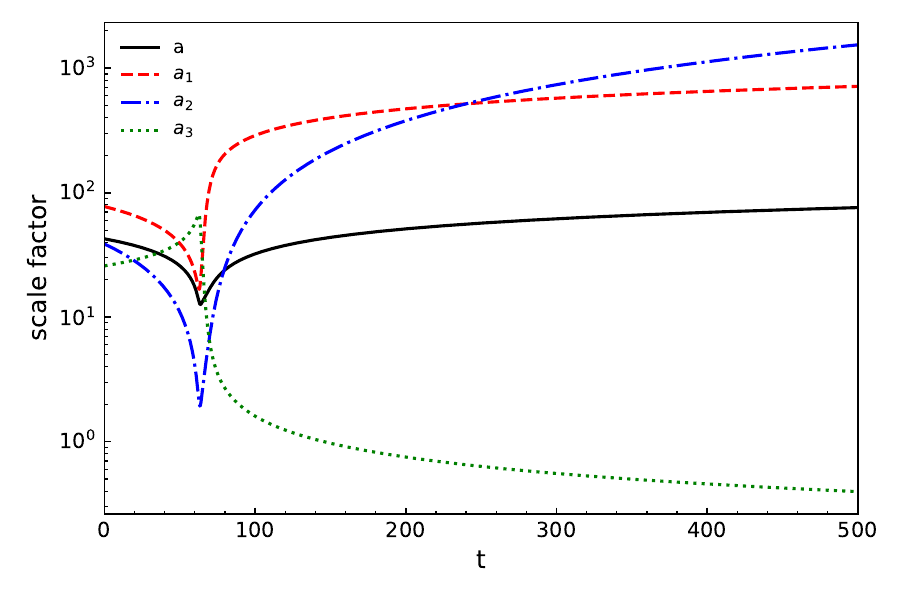}
    \caption{Evolution of directional scale factors and mean scale factor versus time with $p_{1}=1000$, $p_{2}=2000$, $p_{3}=3000$, $c_{1}= -0.13$, $c_{2} = -0.12$ for vacuum Bianchi-I spacetime in the case of $\bar \mu'$ quantization.}
    \label{fig3}
\end{figure}

\begin{figure}[t!]
    \centering
    \includegraphics[scale = 0.6]{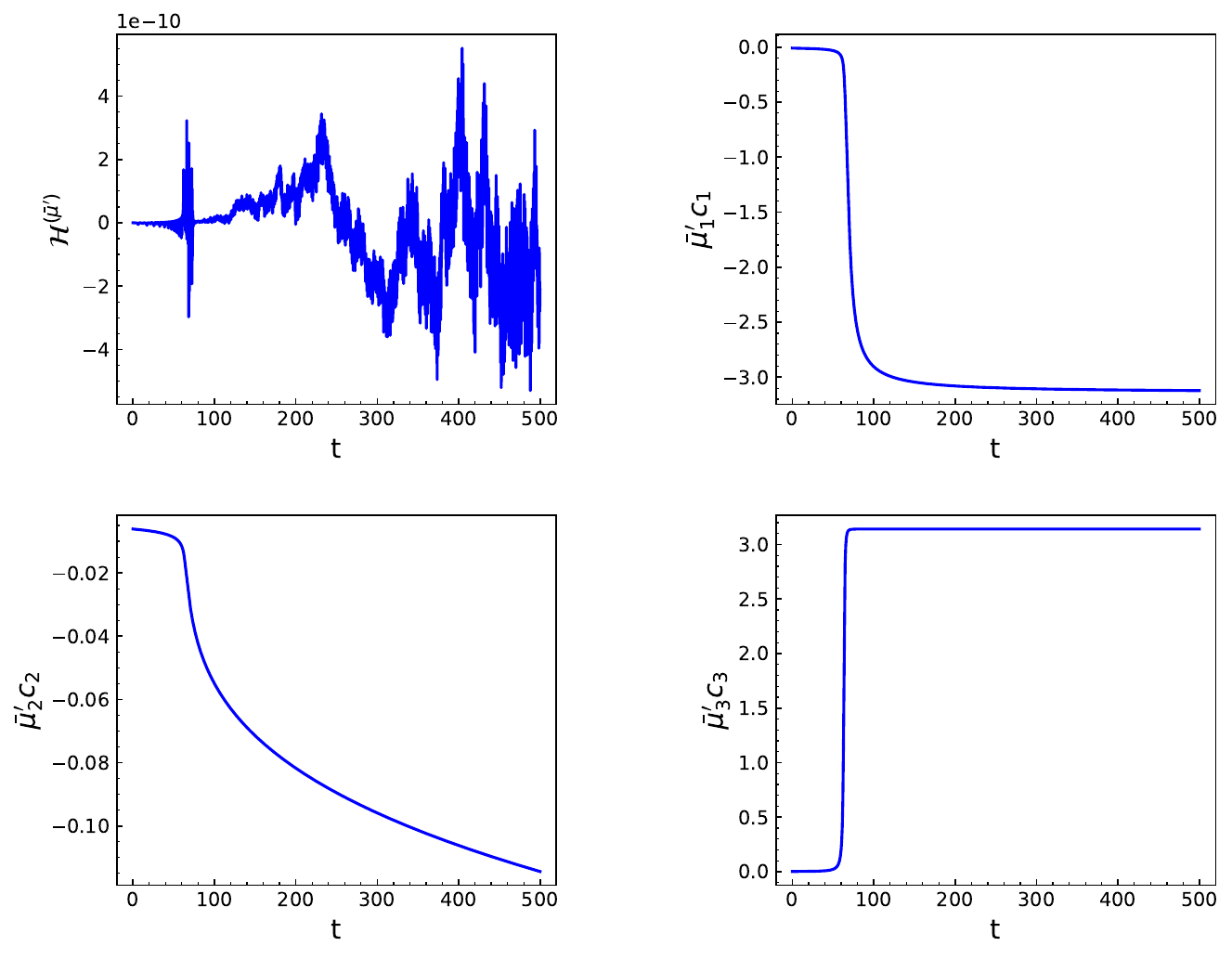}
    \caption{Evolution of Hamiltonian constraint and $\bar\mu_{i}' c_{i}$ versus time with $p_{1}=1000$, $p_{2} =2000$, $p_{3}=3000$, $c_{1}= -0.13$, and $c_{2}= -0.12$ for vacuum Bianchi-I spacetime in the case of $\bar \mu'$ quantization. Unlike the case of $\bar \mu$ quantization, one of the $\bar \mu'_i c_i$ does not reach classical value even long after the bounce.}
    \label{fig4}
\end{figure}

In Fig. \ref{fig1}, we plotted an example evolution with time evolution of all three directional scale factors and also the mean scale factor for $p_{1}=1000$, $p_{2} =2000$, $p_{3}=3000$, $c_{1}= -0.13$, and $c_{2}= -0.12$ for $\bar \mu$ quantization in the case of vacuum Bianchi-I spacetime. As it is obvious, the approach to the singularity is cigar-like in the sense that one of the directional scale factors always contracts and two others expand. In Fig. \ref{fig2}, we  plotted the evolution of $\bar \mu_{i}c_{i}$ versus time,  from the top right, bottom left, and bottom right panels. It is clear that the universe begins in the classical regime while $|\bar \mu_{i} c_{i}|\ll 1$ in the contracting branch and then it becomes classical in the expanding branch, i.e., $|\bar \mu_{i} c_{i}| \sim \pi$. The solution remains reliable as the Hamiltonian constraint is consistently satisfied with high accuracy throughout the universe's time evolution (as shown in the top left panel of Fig. \ref{fig1}). Therefore, the classical cigar-like universe in pre-bounce branch evolves into a classical cigar-like universe in the post-branch and, since the universe becomes classical in the post-bounce branch, the effective dynamics is valid even when one of the directional scale factors becomes small, which happens in the classical regime. These results indicate that in the case of $\bar \mu$ quantization, the universe starts in a large classical regime, bounces due to quantum geometry effects, and then becomes classical at large volumes as expected.

On the other hand, in Fig.\ref{fig3}, we plotted the time evolution of directional scale factors and the mean scale factor in the case of $\bar \mu'$ quantization for the same initial conditions used for $\bar \mu$ quantization. Again, the approach to singularity is cigar-like as it is in a vacuum Bianchi-I spacetime. We then plotted the time evolution of classicality condition for $\bar \mu'$ quantization with the same initial conditions used for $\bar \mu$ quantization in Fig. \ref{fig4}. In this case, although $\bar \mu_{1}' c_{1}$ and $\bar \mu_{3}'c_{3}$ (top right and bottom right panels in Fig. \ref{fig4}) become classical after the bounce, $\bar \mu_{2}' c_{2}$ (bottom left panel in Fig. \ref{fig4}) does not become classical and remains in the quantum regime after the bounce. One can also check that the Hamiltonian constraint is always satisfied with great accuracy in the top left panel of Fig. \ref{fig4}. Hence, we surprisingly found that the $\bar \mu'$ quantization of vacuum Bianchi-I spacetime fails to recover classical GR at a large volume regime, while it is the case for $\bar \mu'$ quantization. Furthermore, given that the universe does not transition to a classical regime in the post-bounce branch, one may also raise questions about the validity of effective dynamics, especially when one of the directional scale factors becomes less than unity, as it is the case here. Note that this precisely the case in one of the quantizations for the Kruskal spacetimes (the Boehmer-Vandersloot quantization) in LQC which is also based on improved dynamics (see Sec IVD of \cite{Ashtekar:2018cay}). However, one may question the robustness of these results, arguing that the phase space of the initial conditions is vast. To address this issue, we show that such quantum behavior results in the violation of conservation of anisotropic shear scalar for a large number of simulations while randomizing the initial conditions in the next section.

\begin{figure}[tbh!]
\centering
    \includegraphics[scale=0.5]{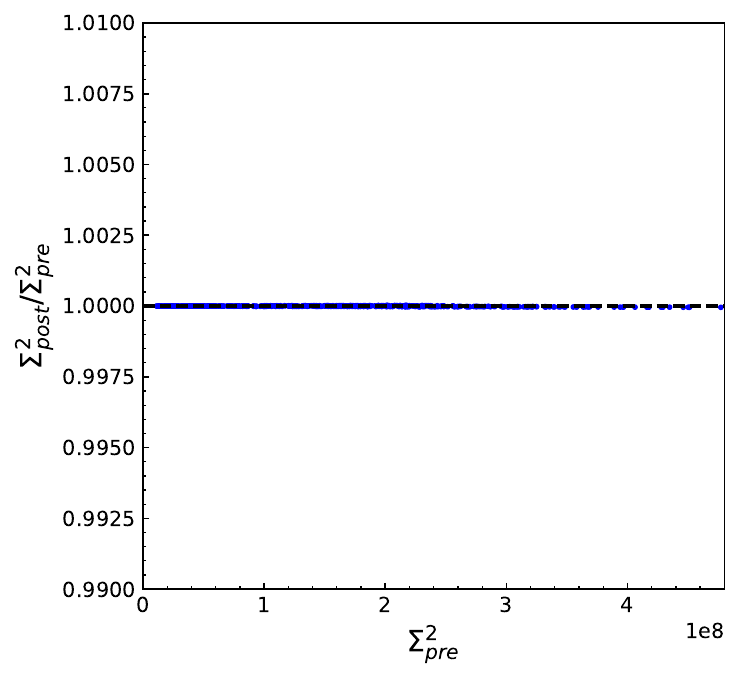} \ \ \ \ \ \ \ \ \ \ \ \ \ \ \  \includegraphics[scale=0.5]{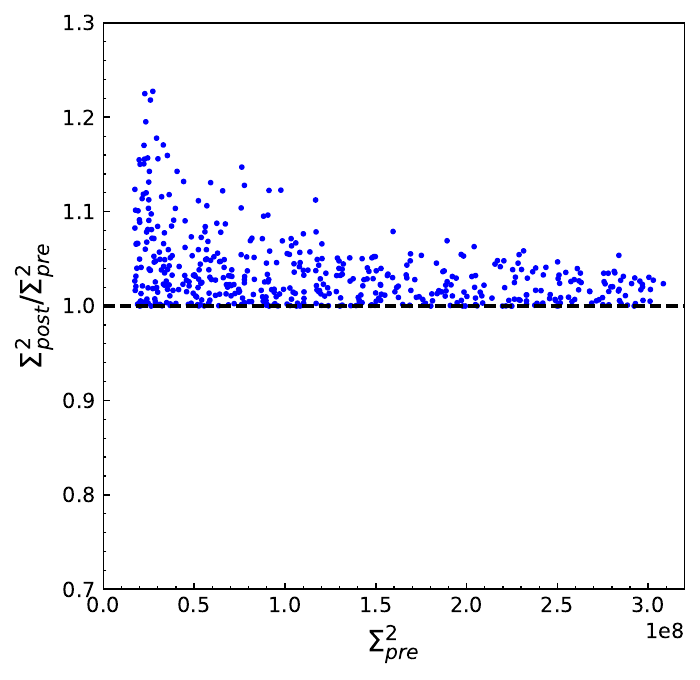}
    \caption{The ratio of post-bounce to pre-bounce anisotropic shear versus pre-bounce anisotropic shear for $\bar \mu$ quantization (left) and $\bar \mu'$ quantization (right).}
    \label{fig5}
\end{figure}

\begin{figure}[t!]
\centering
    \includegraphics[scale=0.52]{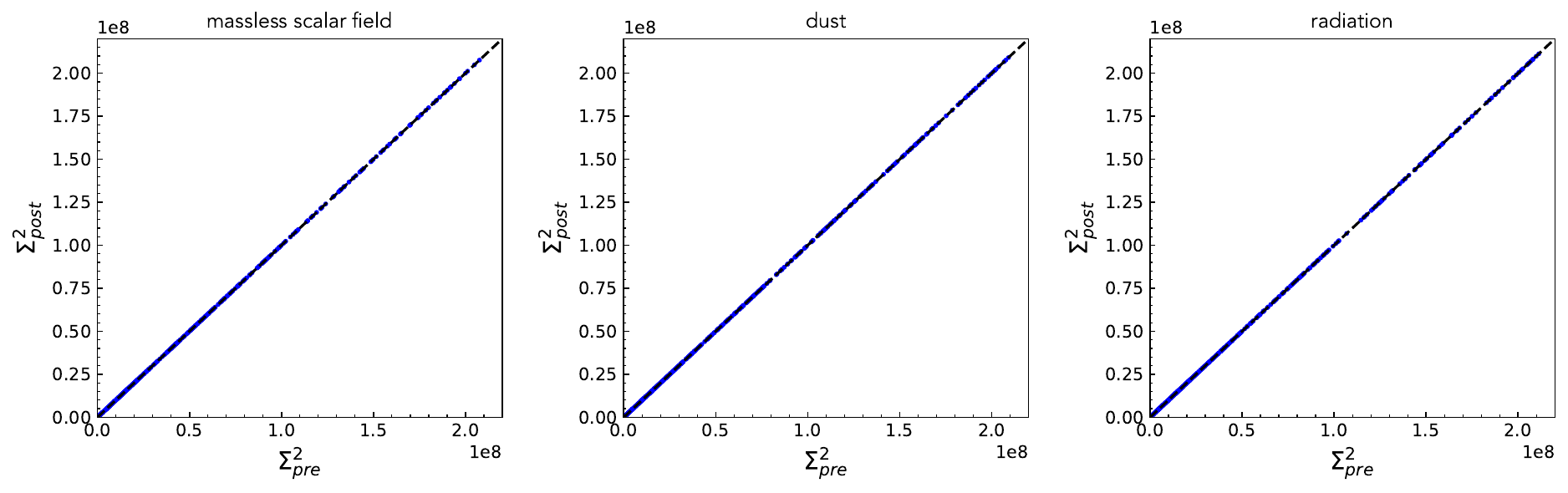}
    \caption{Post-bounce versus pre-bounce anisotropic shear for $\bar \mu$ quantization with different matter contents (massless scalar field, dust, and radiation). We set $p_{1} = 10000$, $p_2 = 20000$, $p_{3} = 30000$ with $c_{1}>0$, $c_{2}>0$ and $c_{3}<0$. The initial value of energy densities are taken to be $\{\rho_{\phi 0}, \rho_{d0}, \rho_{r0}\} = 10^{-6}$. The shear scalar $\Sigma^2$ is preserved across the bounce.}
    \label{fig6}
\end{figure}
\begin{figure}[t!]
\centering
    \includegraphics[scale=0.52]{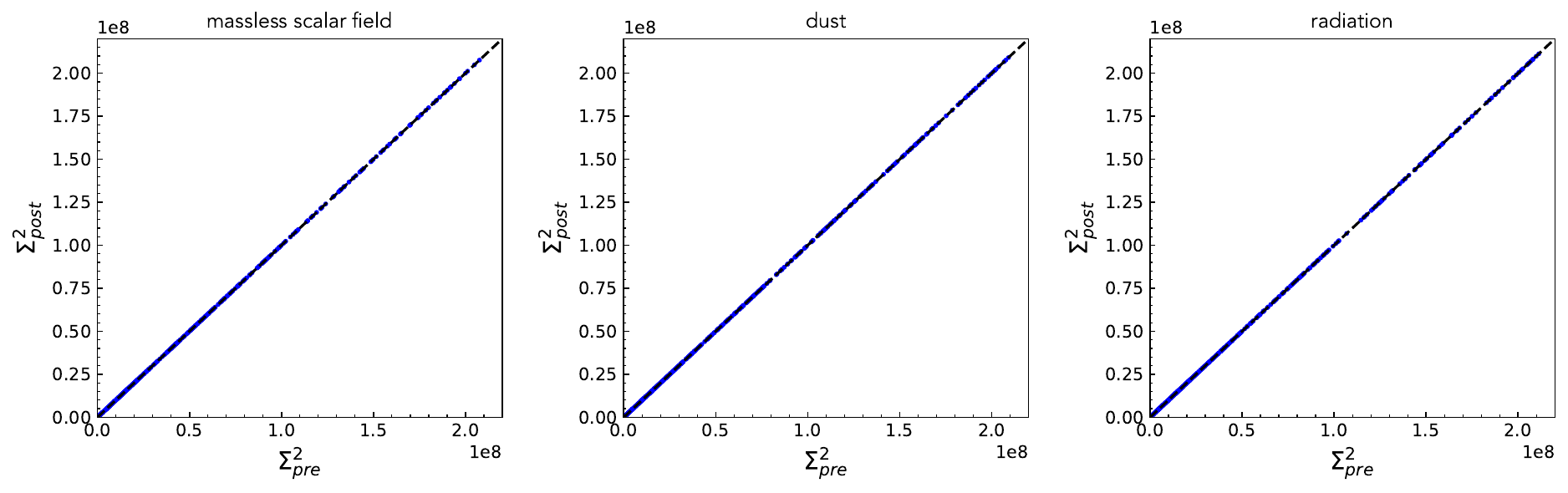}
    \caption{Post-bounce versus pre-bounce anisotropic shear for $\bar \mu$ quantization with different matter contents (massless scalar field, dust, and radiation). We set $p_{1} = 10000$, $p_2 = 20000$, $p_{3} = 30000$ with $c_{1}>0$, $c_{2}>0$ and $c_{3}>0$. The initial value of energy densities are taken to be $\{\rho_{\phi 0}, \rho_{d0}, \rho_{r0}\} = 10^{-6}$.}
    \label{fig7}
\end{figure}

\subsection{Conservation of anisotropic shear}

If the universe becomes classical after the bounce, one expects that the anisotropic shear will be preserved across the bounce, as it is in classical spacetime. However, from the above discussion one may suspect the conservation of anisotropic shear for $\bar \mu'$ quantization since we realize that the universe does not become classical after the bounce. To extract a robust result, we need to compare the pre-bounce to the post-bounce anisotropic shear in the large volume regime for several simulations while sweeping the phase space of the initial conditions. We choose the initial conditions in such a way that the universe starts from a contracting branch at the large volume regime, bounces, and then re-expands in the large volume regime. We fix the initial value of triads in the contracting branch to be at a comparably large volume, then we determine one directional connection using the Hamiltonian constraint, $\mathcal{H}^{(\bar \mu)} =0$ for $\bar \mu$ quantization or $\mathcal{H}^{(\bar \mu')} =0$ for $\bar \mu'$ quantization, and finally randomize two other directional connections while forcing all three directional connections to satisfy classicality conditions, i.e., $|\bar \mu_{i}c_{i}|\ll 1$. To do that, we use \textsc{numbalsoda} package together with the \textsc{numba} package to parallelize and speed up the code so that it could be run on HPC in the case of a large number of simulations. Furthermore, we use the interpolate module in \textsc{scipy} package with the cubic method to find the value of anisotropic shear at the same volume before and after the bounce. Finally, we ran the code for a large number of simulations and excluded those simulations in which the Hamiltonian constraint exceeded $10^{-9}$ during the evolution, and kept the values for pre-bounce and post-bounce anisotropic shear for $500$ simulations.

\begin{figure}
\centering
    \includegraphics[scale=0.52]{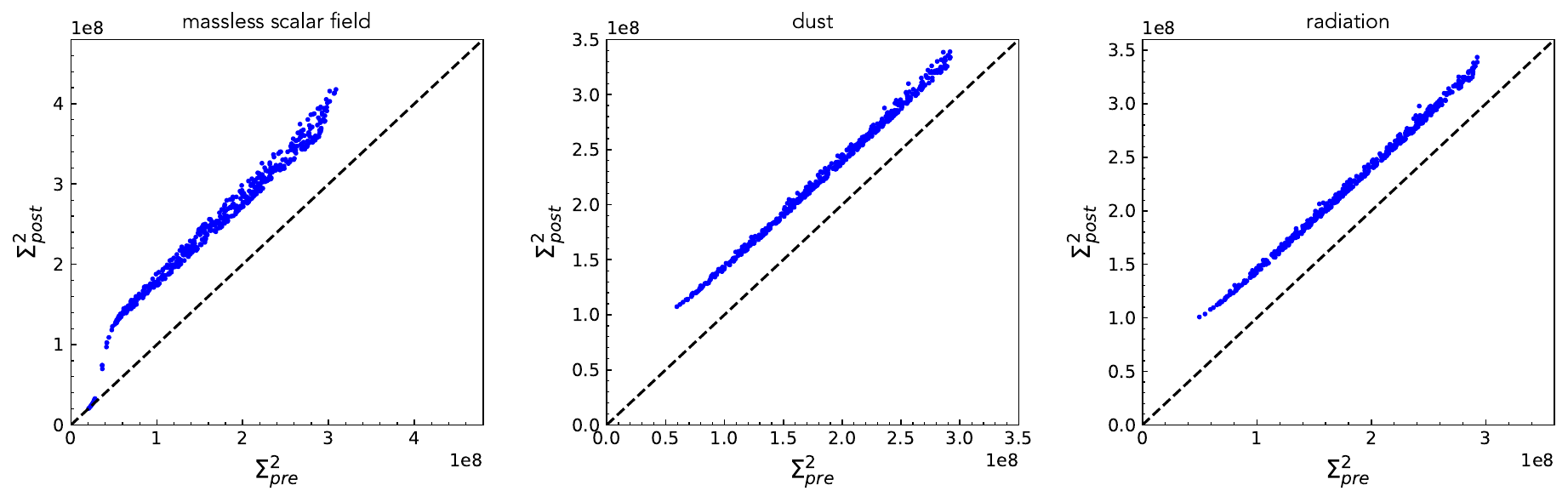}
     \caption{Post-bounce versus pre-bounce anisotropic shear for $\bar \mu'$ quantization with different matter contents (massless scalar field, dust, and radiation). We set $p_{1} = 1000$, $p_2 = 20000$, $p_{3} = 30000$ with $c_{1}<0$, $c_{2}<0$ and $c_{3}>0$. The initial value of energy densities are taken to be $\{\rho_{\phi 0}, \rho_{d0}, \rho_{r0}\} = 10^{-6}$. The shear scalar $\Sigma^2$ is not preserved across the bounce.}
    \label{fig8}
\end{figure}

\begin{figure}
\centering
    \includegraphics[scale=0.52]{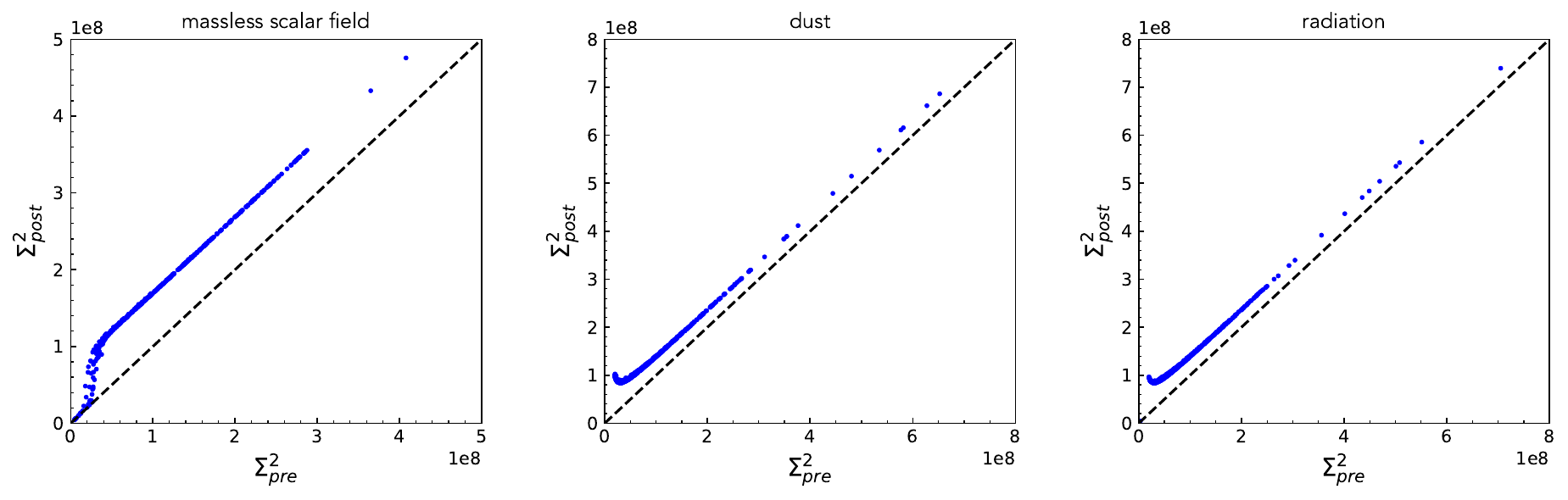}
    \caption{Post-bounce versus pre-bounce anisotropic shear for $\bar \mu'$ quantization with different matter contents (massless scalar field, dust, and radiation). We set $p_{1} = 1000$, $p_2 = 20000$, $p_{3} = 30000$ with $c_{1}<0$, $c_{2}<0$ and $c_{3}<0$. The initial value of energy densities are taken to be $\{\rho_{\phi 0}, \rho_{d0}, \rho_{r0}\} = 10^{-6}$.}
    \label{fig9}
\end{figure}

In Fig. \ref{fig5}, we plotted the ratio of post-bounce to pre-bounce anisotropic shear versus pre-bounce anisotropic shear for  $p_{1} = 10000$, $p_{2} = 20000$ and $p_{3}=30000$ for $\bar \mu$ quantization (left panel) and $\bar \mu'$ quantization (right panel) for the vacuum spacetime. Each dot in the plot corresponds to one simulation, and each plot contains $500$ simulations. From the left panel in Fig. \ref{fig5}, one can see that the anisotropic shear has the same value before and after the bounce in the large volume regime with great accuracy. However, from the right panel in Fig. \ref{fig5}, it is obvious that the post-bounce anisotropic shear has a larger value than the pre-bounce anisotropic shear, even in the large volume regime, indicating the violation of conservation of anisotropic shear in classical Bianchi-I spacetime with $\bar \mu'$ quantization. In order to check the robustness of the results, we ran the code to include isotropic matter contents such as massless sclar field, dust, and radiation. We fixed the energy density of matter  such that $\{\rho_{\phi0}, \rho_{d0}, \rho_{r0}\} = 10^{-6}$ (in Planck units) for all three matter contents. Since these matter components are isotropic, one expects to observe conservation of the anisotropic shear in the classical regime at large volume regime in these cases. Note that in the case of massless scalar field,  the classical singularity  can  either be point-like or cigar-like depending on initial conditions.

We plotted the post-bounce anisotropic shear versus the pre-bounce anisotropic shear for a massless scalar field, dust, and radiation, and for the case that one of connections is positive ($c_{1}<0$, $c_{2}<0$ and $c_{3}>0$) in Fig. \ref{fig6}, and for the case that all connections are negative ($c_{1}<0$, $c_{2}<0$ and $c_{3}<0$) in Fig. \ref{fig7} in the case of $\bar \mu$ quantization. One can see that the anisotropic shear values for all simulations lie on the diagonal, meaning that pre-bounce and post-bounce anisotropic shear have the same value, whereby anisotropic shear is conserved in the $\bar \mu$ quantization of Bianchi-I LQC. On the other hand, we plotted the post-bounce anisotropic shear versus the pre-bounce anisotropic shear again for the massless scalar field, dust, and radiation and for the case that one of connections is positive ($c_{1}<0$, $c_{2}<0$ and $c_{3}>0$) in Fig. \ref{fig8}, and for the case that all connections are negative ($c_{1}<0$, $c_{2}<0$ and $c_{3}<0$) in Fig. \ref{fig9} in the case of $\bar \mu'$ quantization. One can see from these results that the anisotropic shear has a larger value after the bounce compared to the pre-bounce anisotropic shear value. Moreover, as the value of anisotropic shear increases, it seems that the ratio of post-bounce to pre-bounce anisotropic shear reaches a constant value. We should also point out that in the case of a massless scalar field it seems that the anisotropic shear is conserved for a small value of $\Sigma^2$, however, that is because the scale factor dependence of a massless scalar field behaves as anisotropic shear and the universe becomes isotropic due to domination of massless scalar field over anisotropic shear term for the considered initial conditions.

\begin{figure}[t!]
    \centering
    \includegraphics[scale= 0.65]{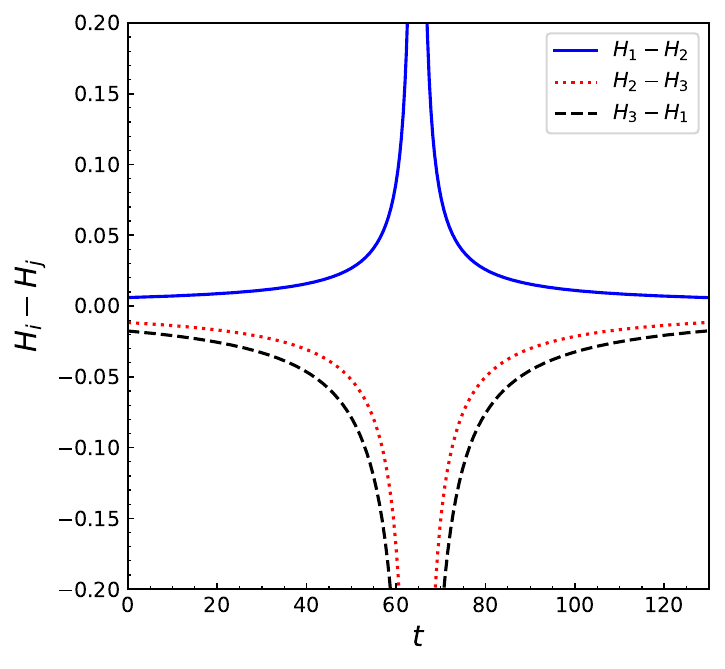} \ \ \ \ \ \ \ \  \ \ \ \includegraphics[scale= 0.65]{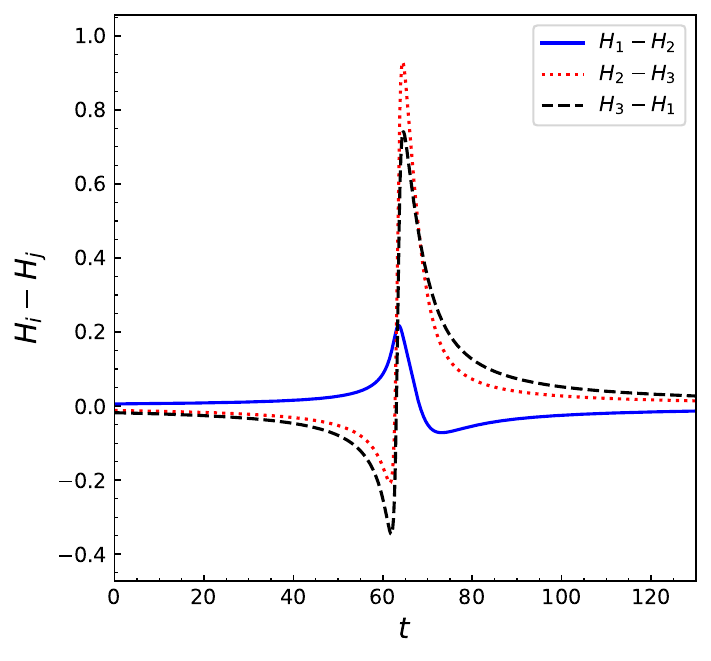}
    \caption{Evolution of difference of directional Hubble parameters, i.e., ($H_{i}-H_{j}$), versus time for $\bar \mu$ quantization (left) and $\bar \mu'$ quantization (right) with $p_{1}=1000$, $p_{2} =2000$, $p_{3}=3000$, $c_{1}= -0.13$, and $c_{2}= -0.12$. Note the contrasting asymmetry for the $\bar \mu'$ prescription.}
    \label{fig10}
\end{figure}
\begin{figure}[t!]
    \centering
    \includegraphics[scale= 0.65]{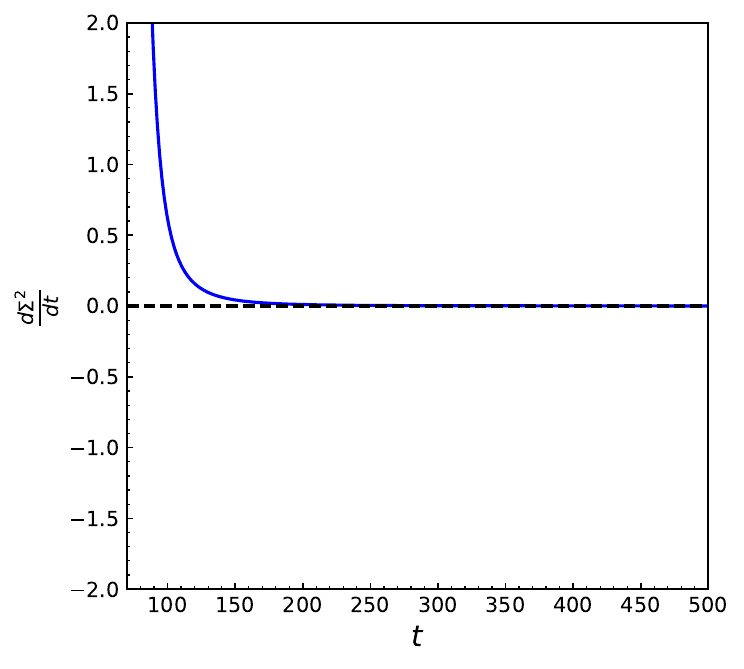} \ \ \ \ \ \ \ \  \ \ \ \includegraphics[scale= 0.65]{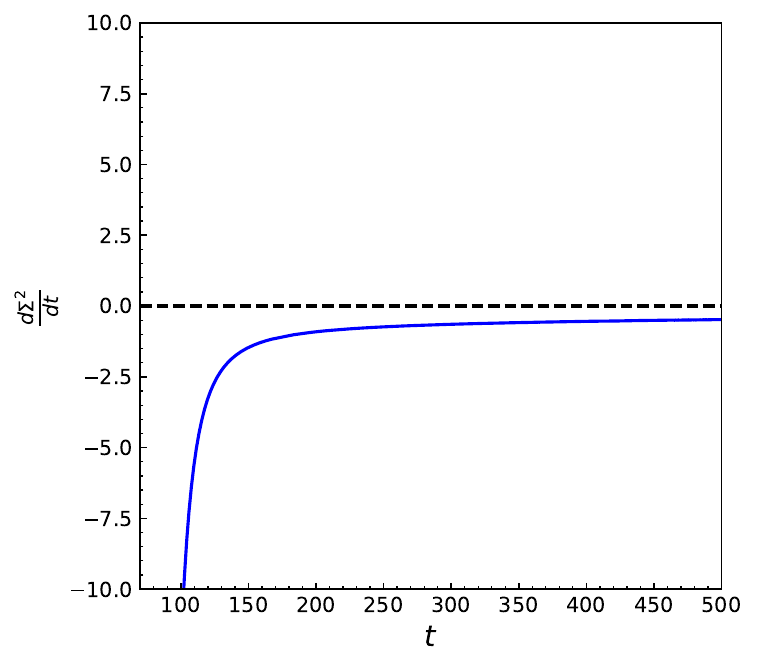}
     \caption{Evolution of $\frac{d\Sigma^2}{dt}$ versus time for $\bar \mu$ quantization (left) and $\bar \mu'$ quantization (right) with $p_{1}=1000$, $p_{2} =2000$, $p_{3}=3000$, $c_{1}= -0.13$, and $c_{2}= -0.12$. The time derivative of shear scalar does not vanish even after a very long time for $\bar \mu'$ prescription. }
    \label{fig11}
\end{figure}

To understand this behavior in detail, we also plotted the time evolution of difference of directional Hubble parameters, i.e., $(H_{i}-H_{j})$, for $p_{1}=1000$, $p_{2}=2000$, $p_{3} = 3000$, $c_{1} = -0.13$ and $c_{2} = -0.12$ for vacuum Bianchi-I spacetime for $\bar \mu$ qunatization (left panel of Fig. \ref{fig10}) and for $\bar \mu'$ quantization (right panel of Fig. \ref{fig10}). From these plots, it is obvious that the difference of directional Hubble parameters before and after the bounce has the same value in the case of $\bar \mu$ quantization, while in the case of $\bar \mu'$ quantization, $(H_{3}-H_{1})$ (black dashed line in the right panel of Fig. \ref{fig10}), has different value in the pre-bounce and post-bounce branches whereby the anisotropic shear is not conserved. Moreover, we also plotted the time evolution of $\frac{d\Sigma^2}{dt}$ for $\bar \mu$ quantization (left panel in Fig.\ref{fig12}) and $\bar \mu'$ quantization (right panel in Fig. \ref{fig12}) from which one can see that $\Sigma^2$ become constant after the bounce in the case of $\bar \mu$ quantization while in the case of $\bar \mu'$, $\Sigma^2$ is not a constant of motion. To conclude, we find that the anisotropic shear is not conserved in the case of $\bar \mu'$ quantization in large volume regimes, where the universe tends to become classical. This violation of anisotropic shear conservation is closely related to the fact that one of the triads remains in the quantum regime and does not become classical even in the large volume regime. This implies that $\bar \mu'$ quantization fails to recover GR in the classical regime, contrary to as was believed in the earlier studies of this model.

\subsection{Non-Classical Cyclic Behavior of $\bar \mu'$ prescription}

\begin{figure}
    \centering
    \includegraphics[scale=0.6]{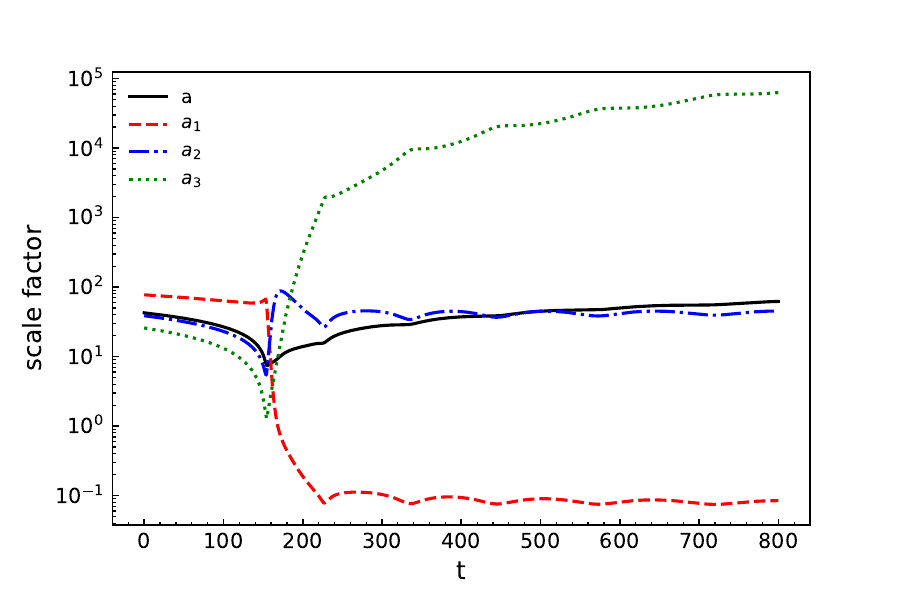}
    \caption{Evolution of directional scale factors and mean scale factor versus time for $p_{1}=1000$, $p_{2}=2000$, $p_{3}=3000$, $c_{1}= c_{2}= - 0.03$, in presence of radiation with $\rho_{r0} = 10^{-6}$ in the $\bar \mu'$ quantization of Bianchi-I spacetime. Non-classical cyclic behavior is observed in directional scale factors.}
    \label{fig12}
\end{figure}

\begin{figure}
    \centering
    \includegraphics[scale=0.6]{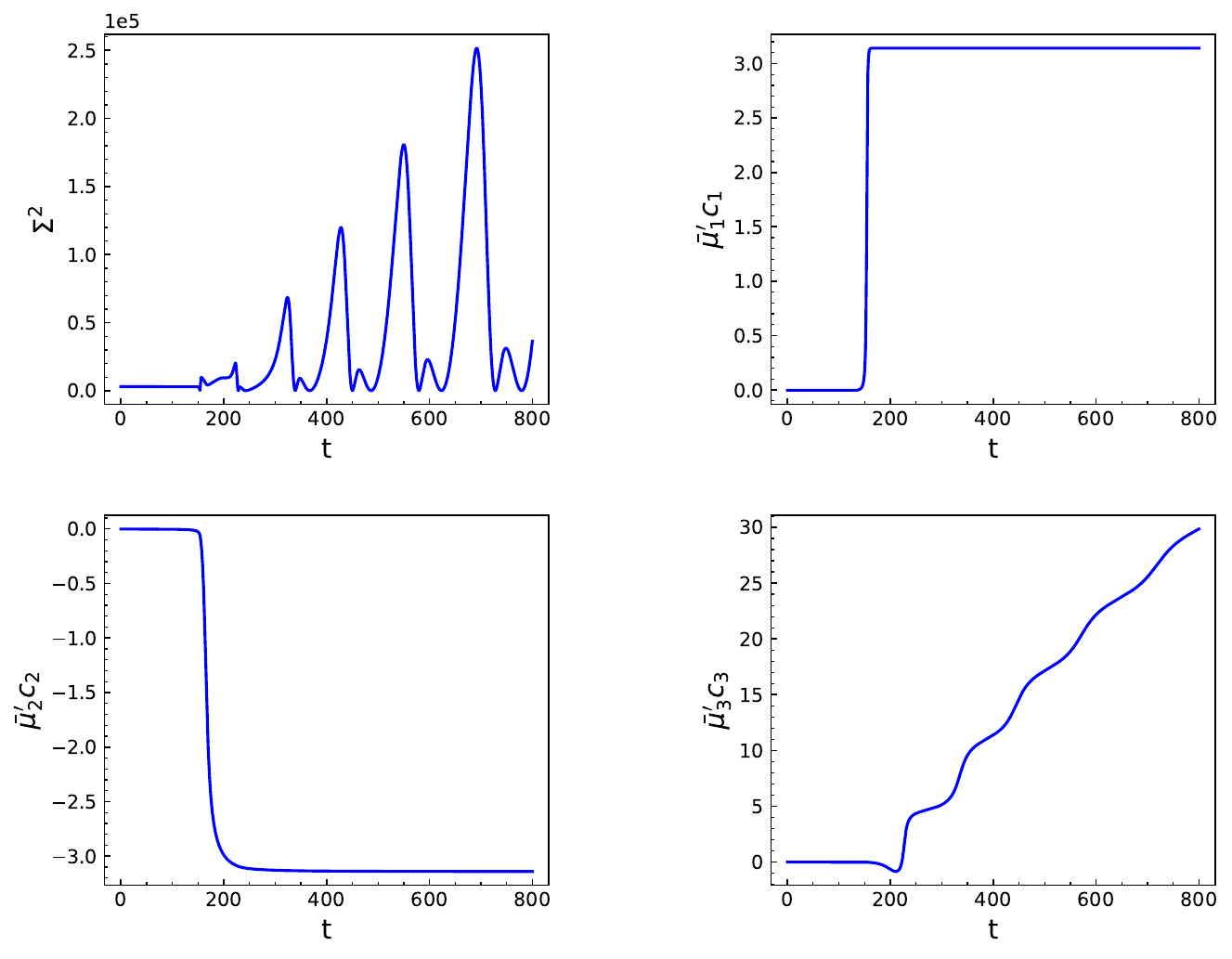}
    \caption{Evolution of anisotropic shear and $\bar \mu_{i}' c_{i}$ versus time for $p_{1}=1000$, $p_{2}=2000$, $p_{3}=3000$, $c_{1} = c_{2} = -0.03$, and $\rho_{r0} = 10^{-6}$ in the $\bar \mu'$ quantization of Bianchi-I spacetime.}
    \label{fig13}
\end{figure}

\begin{figure}
    \centering
    \includegraphics[scale=0.6]{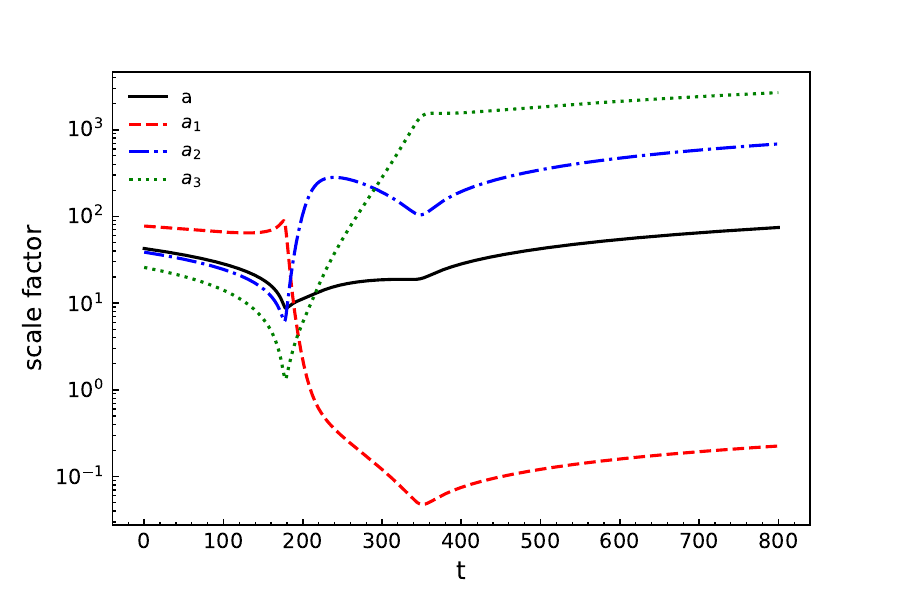}
    \caption{Evolution of directional scale factors and mean scale factor versus time for $p_{1}=1000$, $p_{2}=2000$, $p_{3}=3000$, $c_{1}= c_{2}= - 0.03$, in presence of dust with $\rho_{m0} = 10^{-6}$ in the $\bar \mu'$ quantization of Bianchi-I spacetime.}
    \label{fig14}
\end{figure}
\begin{figure}
    \centering
    \includegraphics[scale=0.6]{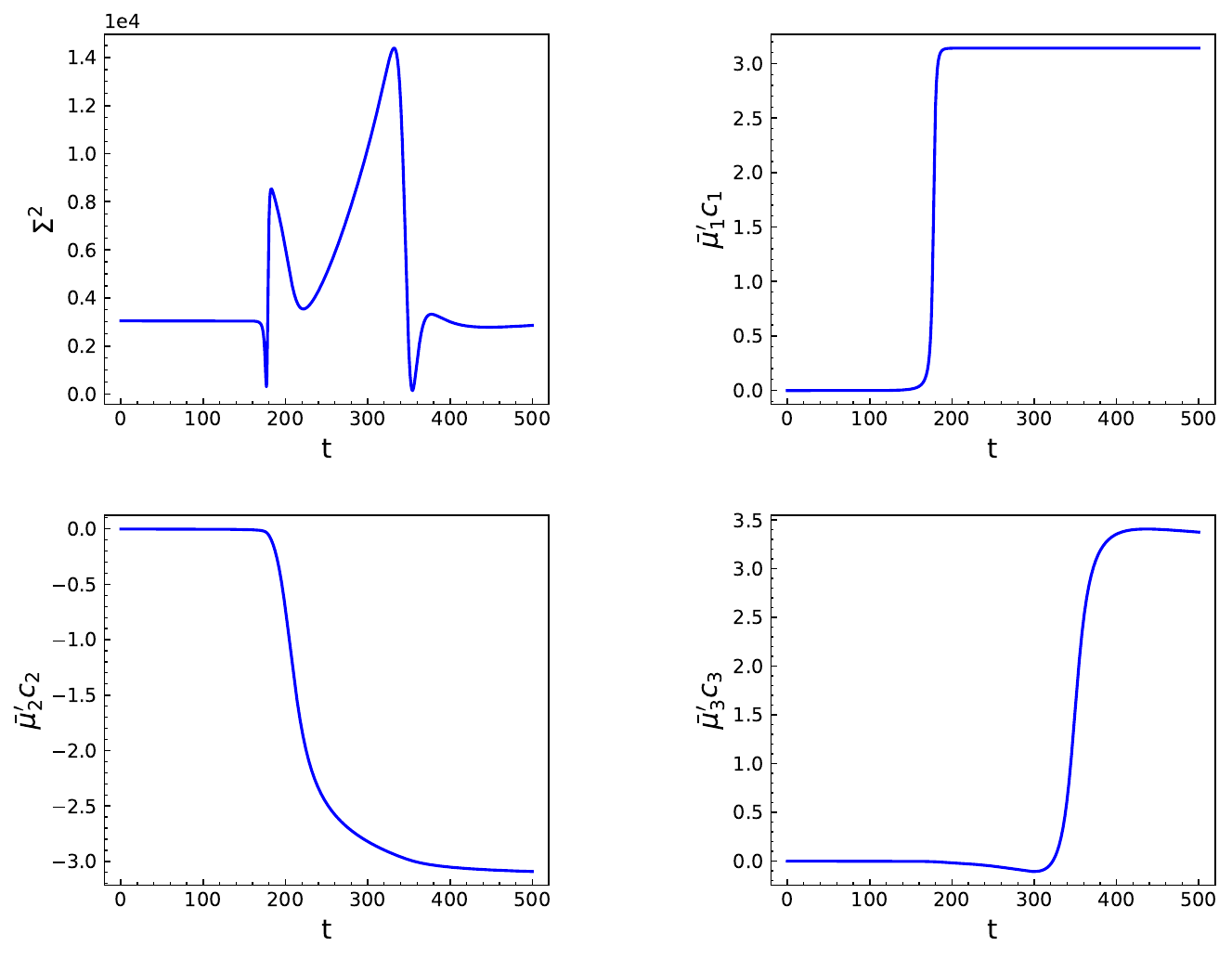}
    \caption{Evolution of anisotropic shear and $\bar \mu_{i}' c_{i}$ versus time for $p_{1}=1000$, $p_{2}=2000$, $p_{3}=3000$, $c_{1} = c_{2} = -0.03$, and $\rho_{m0} = 10^{-6}$ in the $\bar \mu'$ quantization of Bianchi-I spacetime.}
    \label{fig15}
\end{figure}

In the previous section, we surprisingly found that the anisotropic shear is not conserved in the case of $\bar \mu'$ quantization of Bianchi I LQC, which was related to the fact that the universe does not become classical after the bounce. To better understand such an unexpected result, we did extensive numerical analysis for different initial conditions while tracking all relevant quantities and adding different matter content such as a massless scalar field, dust, and radiation. Adding dust and radiation, we realized that, for particular initial conditions, the loop quantum Bianchi-I universe with $\bar \mu'$ prescription goes through some cyclic evolution, which is not expected since radiation and dust cannot produce a recollpsing turnaround point in a classical Bianchi-I spacetime leading into a cyclic universe. As illustrative examples, we plotted the evolution of scale factors and the mean scale factor in Fig. \ref{fig12} for $p_{1}=1000$, $p_{2} =2000$, $p_{3} = 3000$, $c_{1}= c_{2} =-0.03$, and $\rho_{r} = 10^{-6}$ for radiation matter content and in Fig. \ref{fig14} with $\rho_{m0} = 10^{-6}$ for dust matter content. One can see that the universe starts from a contracting branch, bounces back, and then goes through cyclic evolution after the bounce. In Fig. \ref{fig13}, we plotted the time evolution of anisotropic shear and $\bar \mu'_{i} c_{i}$ for the same initial conditions used in Fig. \ref{fig12}. From the top left panel in Fig. \ref{fig13}, it is obvious that the anisotropic shear peaks several times after the bounce, since it goes through several bounces while the maximum anisotropic shear increases for the next bounce. From the top right and bottom left panels in Fig. \ref{fig13}, one can see that the $\bar \mu'_{1}c_{1}$ and $\bar \mu'_{2} c_{2}$ start at classical regime, i.e., $|\bar \mu'_{i} c_{i}| \ll 1$ and become classical after the bounce, i.e., $|\bar \mu'_{i} c_{i}| \sim \pi$. However, from the bottom right panel in Fig. \ref{fig13}, one can find that although $\mu'_{3} c_{3}$ was classical before the bounce at large volume, it did not become classical after the bounce at large volume. Similar behaviors are also observed in the case of dust matter content as it is illustrated in Fig. \ref{fig15}. However, we could not observe such cyclic behavior by adding a massless scalar field, which means that the cyclic behavior is sensitive to the equation of state. We believe that this cyclic behavior in the presence of dust and radiation is tied  to the fact that one of the triads remains in the quantum regime during the evolution of the universe. Therefore,  the $\bar \mu'$ quantization of Bianchi-I spacetime does not recover the GR limit since not only the anisotropic shear is not conserved in the classical regime, but  there is also some unexpected non-classical cyclic behavior.


\section{Conclusions}\label{section V}

To quantize homogeneous Bianchi-I spacetime using techniques from LQG, one faces the same quantization ambiguities as in any other quantum theory. Such quantization ambiguities lead to two different effective descriptions for Bianchi-I LQC; first, $\bar \mu$ quantization, which is consistently written down for non-compact as well as compact manifolds, and in which the energy density and anisotropic shear are universally bounded, second, $\bar \mu'$ quantization, which is non-singular and consistent with compact spatial topology. It has been so far believed that both of these quantizations recover GR in large volume regimes where the universe becomes classical. This means that an effective description of Bianchi-I spacetime should recover the properties of the classical Bianchi-I spacetime in the classical regime. One of the most important properties of classical Bianchi-I spacetime is that an anisotropic shear scalar is a constant of motion. However, this quantity is no longer conserved during the quantum evolution of the universe near the bounce. The presence of the bounce indicates that the universe can start at a large classical regime, contract, bounce back, and expand to a large classical regime again. Therefore, one expects that the anisotropic shear to be conserved when it is compared in a large volume regime before and after the bounce, where the universe is in a classical regime. In other words, one would expect that anisotropic shear would be conserved across the bounce. This issue was analytically investigated in Ref. \cite{Chiou:2007sp} in which it was shown that assuming a large volume limit or classicality condition, i.e., $|\bar \mu_{i} c_{i}| \ll 1$, the effective dynamics recover GR and anisotropic shear is conserved across the bounce. However, since Hamilton's equations are quite complicated given the LQG effects, assuming the classicality condition to hold requires further evidence. In fact, it is possible that the dynamical laws do not allow the universe to become classical after the bounce, which we show is the case for the $\bar \mu'$ scheme.

To address the question whether loop quantum cosmological Bianchi-I spacetime becomes classical, we revisit the homogeneous Bianchi-I model at the level of effective dynamics for both $\bar \mu$ and $\bar \mu'$ quantization. We numerically solved six coupled first-order Hamilton's equations by fixing the triads at a comparably large volume and fixing one of the directional connection components using Hamiltonian constraint while randomizing the other two directional connection components, forcing all three to satisfy classicality conditions. We found that the universe starts at the classical regime, bounces, and then expands back to the classical regime in the case of $\bar \mu$ quantization. Surprisingly, we find that although the universe is in the classical regime before the bounce, it does not become classical in the expanding branch, since one of the triads and associated trigonometric term in the Hamiltonian constraint always remains in the quantum regime in the case of $\bar \mu'$ quantization. This implies that the classicality condition is not satisfied when the universe is at a large volume after the bounce. The analysis done in Ref. \cite{Chiou:2007sp} is no longer valid, and one may suspect the conservation of anisotropic shear across the bounce. To address this issue, we compared the anisotropic shear before and after the bounce at the same volume for a large number of simulations. In the case of $\bar \mu$ quantization, we found that the anisotropic shear is conserved, as expected, with great accuracy before and after the bounce when the universe becomes classical at large volume regime. On the other hand,  the anisotropic shear is not conserved across the bounce for $\bar \mu'$ quantization. In fact, such a violation of the conservation of anisotropic shear is closely related to the fact that the universe does not become classical after the bounce. In other words, since the universe is still in the quantum regime, it does not recover the classical properties of Bianchi-I spacetime, i.e., a conserved anisotropic shear. To check the robustness of this result, we included a massless scalar field, dust, and radiation matter fields, which are isotropic matter contents, and one would expect that the anisotropic shear would be conserved across the bounce in these cases indeed. Again, we find that the anisotropic shear is not conserved across the bounce in all three cases for $\bar \mu'$ quantization. Doing an exhaustive numerical analysis, we find that for particular initial conditions, including dust or radiation, the loop quantized Bianchi-I spacetime with $\bar \mu'$ prescription exhibits some unexpected cyclic behavior. Although we could not observe such cyclic behavior by adding massless scalar field which means that the cyclic behavior is sensitive to the equation of state. We believe that such peculiar cyclic behavior is again due to the fact that the universe remains in the quantum regime after the bounce. Based on these results, we conclude that the $\bar \mu'$ quantization of Bianchi-I spacetime does not become classical at large volume regime. Finally, only the $\bar \mu$ is a unique loop quantization of Bianchi-I spacetime which is physically viable, for both compact and non-compact spatial manifolds.

\begin{acknowledgements}
This work is supported by the NSF grant PHY-2110207. Authors acknowledge the support of HPC resources at LSU.
\end{acknowledgements}


\begin{thebibliography}{9}

\bibitem{Geroch:1968ut}
R.~P.~Geroch,
 ``What is a singularity in general relativity?,''
Annals Phys. \textbf{48}, 526-540 (1968).

\bibitem{Hawking:1970zqf}
S.~W.~Hawking and R.~Penrose,
``The Singularities of gravitational collapse and cosmology,''
Proc. Roy. Soc. Lond. A \textbf{314}, 529-548 (1970).

\bibitem{Borde:1993xh}
A.~Borde and A.~Vilenkin,
``Eternal inflation and the initial singularity,''
Phys. Rev. Lett. \textbf{72}, 3305-3309 (1994)
[arXiv:gr-qc/9312022 [gr-qc]].

\bibitem{Borde:2001nh}
A.~Borde, A.~H.~Guth and A.~Vilenkin,
``Inflationary space-times are incompletein past directions,''
Phys. Rev. Lett. \textbf{90}, 151301 (2003)
[arXiv:gr-qc/0110012 [gr-qc]].

\bibitem{Ashtekar:2011ni}
A.~Ashtekar and P.~Singh,
``Loop Quantum Cosmology: A Status Report,''
Class. Quant. Grav. \textbf{28}, 213001 (2011)
[arXiv:1108.0893 [gr-qc]].

\bibitem{Ashtekar:2006rx}
A.~Ashtekar, T.~Pawlowski and P.~Singh,
``Quantum nature of the big bang,''
Phys. Rev. Lett. \textbf{96}, 141301 (2006)
[arXiv:gr-qc/0602086 [gr-qc]].

\bibitem{Ashtekar:2006uz}
A.~Ashtekar, T.~Pawlowski and P.~Singh,
``Quantum Nature of the Big Bang: An Analytical and Numerical Investigation. I.,''
Phys. Rev. D \textbf{73}, 124038 (2006)
[arXiv:gr-qc/0604013 [gr-qc]].

\bibitem{Ashtekar:2006wn}
A.~Ashtekar, T.~Pawlowski and P.~Singh,
``Quantum Nature of the Big Bang: Improved dynamics,''
Phys. Rev. D \textbf{74}, 084003 (2006)
[arXiv:gr-qc/0607039 [gr-qc]].

\bibitem{Ashtekar:2007em}
A.~Ashtekar, A.~Corichi and P.~Singh,
``Robustness of key features of loop quantum cosmology,''
Phys. Rev. D \textbf{77}, 024046 (2008)
[arXiv:0710.3565 [gr-qc]].

\bibitem{Craig:2013mga}
D.~A.~Craig and P.~Singh,
``Consistent probabilities in loop quantum cosmology,''
Class. Quant. Grav. \textbf{30}, 205008 (2013)
[arXiv:1306.6142 [gr-qc]].


\bibitem{Szulc:2006ep}
L.~Szulc, W.~Kaminski and J.~Lewandowski,
``Closed FRW model in Loop Quantum Cosmology,''
Class. Quant. Grav. \textbf{24}, 2621-2636 (2007)
[arXiv:gr-qc/0612101 [gr-qc]].

\bibitem{Ashtekar:2006es}
A.~Ashtekar, T.~Pawlowski, P.~Singh and K.~Vandersloot,
``Loop quantum cosmology of k=1 FRW models,''
Phys. Rev. D \textbf{75}, 024035 (2007)
[arXiv:gr-qc/0612104 [gr-qc]].


\bibitem{Vandersloot:2006ws}
K.~Vandersloot,
``Loop quantum cosmology and the k = - 1 RW model,''
Phys. Rev. D \textbf{75}, 023523 (2007)
[arXiv:gr-qc/0612070 [gr-qc]].

\bibitem{Szulc:2007uk}
L.~Szulc,
``Open FRW model in Loop Quantum Cosmology,''
Class. Quant. Grav. \textbf{24}, 6191-6200 (2007)
[arXiv:0707.1816 [gr-qc]].

\bibitem{Giesel:2020raf}
K.~Giesel, B.~F.~Li and P.~Singh,
``Towards a reduced phase space quantization in loop quantum cosmology with an inflationary potential,''
Phys. Rev. D \textbf{102}, no.12, 126024 (2020)
[arXiv:2007.06597 [gr-qc]].

\bibitem{Ashtekar:2009vc}
A.~Ashtekar and E.~Wilson-Ewing,
``Loop quantum cosmology of Bianchi-I models,''
Phys. Rev. D \textbf{79}, 083535 (2009)
[arXiv:0903.3397 [gr-qc]].

\bibitem{Ashtekar:2009um}
A.~Ashtekar and E.~Wilson-Ewing,
``Loop quantum cosmology of Bianchi type II models,''
Phys. Rev. D \textbf{80}, 123532 (2009)
[arXiv:0910.1278 [gr-qc]].

\bibitem{Wilson-Ewing:2010lkm}
E.~Wilson-Ewing,
``Loop quantum cosmology of Bianchi type IX models,''
Phys. Rev. D \textbf{82}, 043508 (2010)
[arXiv:1005.5565 [gr-qc]].

\bibitem{Garay:2010sk}
L.~J.~Garay, M.~Martin-Benito and G.~A.~Mena Marugan,
``Inhomogeneous Loop Quantum Cosmology: Hybrid Quantization of the Gowdy Model,''
Phys. Rev. D \textbf{82}, 044048 (2010)
[arXiv:1005.5654 [gr-qc]].

\bibitem{Agullo:2016tjh}
I.~Agullo and P.~Singh,
``Loop Quantum Cosmology,''
[arXiv:1612.01236 [gr-qc]].

\bibitem{Li:2021mop}
B.~F.~Li, P.~Singh and A.~Wang,
``Phenomenological implications of modified loop cosmologies: an overview,''
Front. Astron. Space Sci. \textbf{8}, 701417 (2021)
[arXiv:2105.14067 [gr-qc]].

\bibitem{Li:2023dwy}
B.~F.~Li and P.~Singh,
``Loop Quantum Cosmology: Physics of Singularity Resolution and its Implications,''
[arXiv:2304.05426 [gr-qc]].

\bibitem{Gupt:2011jh}
B.~Gupt and P.~Singh,
``Contrasting features of anisotropic loop quantum cosmologies: The Role of spatial curvature,''
Phys. Rev. D \textbf{85}, 044011 (2012)
[arXiv:1109.6636 [gr-qc]].

\bibitem{Corichi:2008zb}
A.~Corichi and P.~Singh,
``Is loop quantization in cosmology unique?,''
Phys. Rev. D \textbf{78}, 024034 (2008)
[arXiv:0805.0136 [gr-qc]].

\bibitem{Singh:2013ava}
P.~Singh and E.~Wilson-Ewing,
``Quantization ambiguities and bounds on geometric scalars in anisotropic loop quantum cosmology,''
Class. Quant. Grav. \textbf{31}, 035010 (2014)
[arXiv:1310.6728 [gr-qc]].

\bibitem{Singh:2010qa}
P.~Singh and F.~Vidotto,
``Exotic singularities and spatially curved Loop Quantum Cosmology,''
Phys. Rev. D \textbf{83}, 064027 (2011)
[arXiv:1012.1307 [gr-qc]].

\bibitem{Saini:2018tto}
S.~Saini and P.~Singh,
``Generic absence of strong singularities and geodesic completeness in modified loop quantum cosmologies,''
Class. Quant. Grav. \textbf{36}, no.10, 105014 (2019)
[arXiv:1812.08937 [gr-qc]].

\bibitem{Singh:2011gp}
P.~Singh,
``Curvature invariants, geodesics and the strength of singularities in Bianchi-I loop quantum cosmology,''
Phys. Rev. D \textbf{85}, 104011 (2012)
[arXiv:1112.6391 [gr-qc]].

\bibitem{Saini:2017ipg}
S.~Saini and P.~Singh,
``Resolution of strong singularities and geodesic completeness in loop quantum Bianchi-II spacetimes,''
Class. Quant. Grav. \textbf{34}, no.23, 235006 (2017)
[arXiv:1707.08556 [gr-qc]].

\bibitem{Saini:2017ggt}
S.~Saini and P.~Singh,
``Generic absence of strong singularities in loop quantum Bianchi-IX spacetimes,''
Class. Quant. Grav. \textbf{35}, no.6, 065014 (2018)
[arXiv:1712.09474 [gr-qc]].

\bibitem{Saini:2016vgo}
S.~Saini and P.~Singh,
``Geodesic completeness and the lack of strong singularities in effective loop quantum Kantowski\textendash{}Sachs spacetime,''
Class. Quant. Grav. \textbf{33}, no.24, 245019 (2016)
[arXiv:1606.04932 [gr-qc]].

\bibitem{Ashtekar:2023cod}
A.~Ashtekar, J.~Olmedo and P.~Singh,
``Regular black holes from Loop Quantum Gravity,'' a
Invited Chapter for the book Regular Black Holes: Towards a New Paradigm of Gravitational Collapse, Ed. C. Bambi, Springer Singapore (2023) arXiv:2301.01309 [gr-qc]

\bibitem{Corichi:2009pp}
A.~Corichi and P.~Singh,
``A Geometric perspective on singularity resolution and uniqueness in loop quantum cosmology,''
Phys. Rev. D \textbf{80}, 044024 (2009)
[arXiv:0905.4949 [gr-qc]].




\bibitem{Chiou:2006qq}
D.~W.~Chiou,
``Loop Quantum Cosmology in Bianchi Type I Models: Analytical Investigation,''
Phys. Rev. D \textbf{75}, 024029 (2007)
[arXiv:gr-qc/0609029 [gr-qc]].

\bibitem{Chiou:2007sp}
D.~W.~Chiou and K.~Vandersloot,
``The Behavior of non-linear anisotropies in bouncing Bianchi-I models of loop quantum cosmology,''
Phys. Rev. D \textbf{76}, 084015 (2007)
[arXiv:0707.2548 [gr-qc]].


\bibitem{Martin-Benito:2008dfr}
M.~Martin-Benito, G.~A.~Mena Marugan and T.~Pawlowski,
``Loop Quantization of Vacuum Bianchi I Cosmology,''
Phys. Rev. D \textbf{78}, 064008 (2008)
[arXiv:0804.3157 [gr-qc]].

\bibitem{Martin-Benito:2009xaf}
M.~Martin-Benito, G.~A.~M.~Marugan and T.~Pawlowski,
``Physical evolution in Loop Quantum Cosmology: The Example of vacuum Bianchi I,''
Phys. Rev. D \textbf{80}, 084038 (2009)
[arXiv:0906.3751 [gr-qc]].


\bibitem{Tsagas:2007yx}
C.~G.~Tsagas, A.~Challinor and R.~Maartens,
``Relativistic cosmology and large-scale structure,''
Phys. Rept. \textbf{465}, 61-147 (2008)
[arXiv:0705.4397 [astro-ph]].



\bibitem{Linde:1981mu}
A.~D.~Linde,
``A New Inflationary Universe Scenario: A Possible Solution of the Horizon, Flatness, Homogeneity, Isotropy and Primordial Monopole Problems,''
Phys. Lett. B \textbf{108}, 389-393 (1982).

\bibitem{Guth:1980zm}
A.~H.~Guth,
``The Inflationary Universe: A Possible Solution to the Horizon and Flatness Problems,''
Phys. Rev. D \textbf{23}, 347-356 (1981).

\bibitem{Bardeen:1983qw}
J.~M.~Bardeen, P.~J.~Steinhardt and M.~S.~Turner,
``Spontaneous Creation of Almost Scale - Free Density Perturbations in an Inflationary Universe,''
Phys. Rev. D \textbf{28}, 679 (1983).


\bibitem{Steinhardt:2002ih}
P.~J.~Steinhardt, N.~Turok and N.~Turok,
``A Cyclic model of the universe,''
Science \textbf{296}, 1436-1439 (2002)
[arXiv:hep-th/0111030 [hep-th]].

\bibitem{Buchbinder:2007ad}
E.~I.~Buchbinder, J.~Khoury and B.~A.~Ovrut,
``New Ekpyrotic cosmology,''
Phys. Rev. D \textbf{76}, 123503 (2007)
[arXiv:hep-th/0702154 [hep-th]].

\bibitem{Gupt:2012vi}
B.~Gupt and P.~Singh,
``Quantum gravitational Kasner transitions in Bianchi-I spacetime,''
Phys. Rev. D \textbf{86}, 024034 (2012)
[arXiv:1205.6763 [gr-qc]].





\bibitem{Diener:2013uka}
P.~Diener, B.~Gupt and P.~Singh,
``Chimera: A hybrid approach to numerical loop quantum cosmology,''
Class. Quant. Grav. \textbf{31}, 025013 (2014)
[arXiv:1310.4795 [gr-qc]].

\bibitem{Diener:2014mia}
P.~Diener, B.~Gupt and P.~Singh,
``Numerical simulations of a loop quantum cosmos: robustness of the quantum bounce and the validity of effective dynamics,''
Class. Quant. Grav. \textbf{31}, 105015 (2014)
[arXiv:1402.6613 [gr-qc]].

\bibitem{Diener:2014hba}
P.~Diener, B.~Gupt, M.~Megevand and P.~Singh,
``Numerical evolution of squeezed and non-Gaussian states in loop quantum cosmology,''
Class. Quant. Grav. \textbf{31}, 165006 (2014)
[arXiv:1406.1486 [gr-qc]].

\bibitem{Diener:2017lde}
P.~Diener, A.~Joe, M.~Megevand and P.~Singh,
``Numerical simulations of loop quantum Bianchi-I spacetimes,''
Class. Quant. Grav. \textbf{34}, no.9, 094004 (2017)
[arXiv:1701.05824 [gr-qc]].

\bibitem{Motaharfar:2022pjp}
M.~Motaharfar and P.~Singh,
``Tunneling wave function proposal with loop quantum geometry effects,''
Phys. Rev. D \textbf{107}, no.6, 066026 (2023)
[arXiv:2212.14065 [gr-qc]]

\bibitem{Motaharfar:2023gpp}
M.~Motaharfar and P.~Singh,
``Quantum Gravitational Non-Singular Tunneling Wavefunction Proposal \textdagger{},''
Phys. Sci. Forum \textbf{7}, no.1, 44 (2023)
[arXiv:2304.06760 [gr-qc]]


\bibitem{McNamara:2022dmf}
A.~M.~McNamara, S.~Saini and P.~Singh,
``Novel relationship between shear and energy density at the bounce in nonsingular Bianchi-I spacetimes,''
Phys. Rev. D \textbf{107}, no.2, 026003 (2023)
[arXiv:2210.07257 [gr-qc]].

\bibitem{Ashtekar:2018cay}
A.~Ashtekar, J.~Olmedo and P.~Singh,
``Quantum extension of the Kruskal spacetime,''
Phys. Rev. D \textbf{98}, no.12, 126003 (2018)
[arXiv:1806.02406 [gr-qc]]

\end{thebibliography}
\end{document}